\documentclass[fleqn,usenatbib]{mnras}

\usepackage{newtxtext,newtxmath}

\usepackage[T1]{fontenc}

\DeclareRobustCommand{\VAN}[3]{#2}
\let\VANthebibliography\thebibliography
\def\thebibliography{\DeclareRobustCommand{\VAN}[3]{##3}\VANthebibliography}

\usepackage{graphicx}	
\usepackage{amsmath}	

\newcommand\tmtsdata{TMTS-ULC1st~}
\newcommand\tmtslocal{``daily''\ }

\title[Minute-cadence Observations of the TMTS I. ]{ Minute-cadence Observations of the LAMOST Fields with the TMTS:\\ I. Methodology of Detecting Short-period Variables and Results from the first-year Survey }

\author[Lin et al.]{
Jie Lin,$^{1}$\thanks{E-mail: linjie2019@mail.tsinghua.edu.cn}
Xiaofeng Wang,$^{1,2}$\thanks{E-mail: wang\_xf@mail. tsinghua.edu.cn}
Jun Mo,$^{1}$\thanks{E-mail: mojun@mails.tsinghua.edu.cn}
Gaobo Xi,$^{1}$
Jicheng Zhang,$^{3}$
Xiaojun Jiang,$^{4}$
Jianrong Shi,$^{4,5}$ \newauthor
Xiaobin Zhang,$^{4}$
Xiaoming Zhang,$^{4}$
Zixuan Wei,$^{1}$
Limeng Ye,$^{6}$
Chengyuan Wu,$^{1}$
Shengyu Yan,$^{1}$ \newauthor
Zhihao Chen,$^{1}$ 
Wenxiong Li,$^{7}$
Xue Li,$^{1}$ 
Weili Lin,$^{1}$
Han Lin,$^{1}$
Hanna Sai,$^{1}$
Danfeng Xiang$^{1}$ \newauthor
and Xinghan Zhang$^{1}$
\\
$^{1}$Physics Department and Tsinghua Center for Astrophysics, Tsinghua University, Beijing, 100084, People's Republic of China\\
$^{2}$Beijing Planetarium, Beijing Academy of Sciences, 100044, Beijing\\
$^{3}$Department of Astronomy, Beijing Normal University, Beijing, 100875, People's Republic of China\\
$^{4}$National Astronomical Observatories of China, Chinese Academy of Sciences, Tsinghua University, Beijing, 100012, People’s Republic of China\\
$^{5}$School of Astronomy and Space Science, University of Chinese Academy of Sciences, Beijing, 100049, People’s Republic of China\\
$^{6}$School of Physics and Astronomy, Queen Mary University of London, G. O. Jones Building, 327 Mile End Road London, E1 4NS, UK\\
$^{7}$The School of Physics and Astronomy, Tel Aviv University, Tel Aviv 69978, Israel
}

\date{Accepted XXX. Received YYY; in original form ZZZ}

\pubyear{202x}

\begin{document}
\label{firstpage}
\pagerange{\pageref{firstpage}--\pageref{lastpage}}
\maketitle

\begin{abstract}
Tsinghua University-Ma Huateng Telescopes for Survey (TMTS), located at Xinglong Station of NAOC, has a field of view up to 18 deg$^2$. The TMTS has started to monitor the LAMOST sky areas since 2020, with the uninterrupted observations lasting for about 6 hours on average for each sky area and a cadence of about 1 minute.
Here we introduce the data analysis and preliminary scientific results for the first-year observations, which covered 188 LAMOST plates ($\approx 1970~{\rm deg}^2$). These observations have generated over 4.9 million uninterrupted light curves, with at least 100 epochs for each of them. These light curves correspond to 4.26 million Gaia-DR2 sources, among which 285 thousand sources are found to have multi-epoch {spectra} from the LAMOST.
By analysing these light curves with the Lomb–Scargle periodograms, we identify more than  3700 periodic variable star candidates with periods below { $\approx 7.5$ hours}, primarily consisting of eclipsing binaries and $\delta$ Scuti stars. Those short-period binaries will provide important constraints on theories of binary evolution and possible sources for space gravitational wave experiments in the future. 
Moreover, we also identified 42 flare stars by searching rapidly-evolving signals in the light curves. The densely-sampled light curves from the TMTS allow us to better quantify the shapes and durations for these flares.

\end{abstract}




\section{Introduction}

The binaries with orbital periods shorter than a few hours, namely { ultracompact binaries (UCBs)}, play a crucial role in the functional tests of space gravitational wave observatories \citep{Shah+etal+2012}. 
Over the past two years, the Zwicky Transient Facility (ZTF) has discovered a few UCBs with orbital period shorter than 20 minutes through densely sampled photometric measurements \citep{Burdge+etal+2019+Nature,Burdge+etal+2020+systematic,Burdge+etal+2020+8.8min}, these binaries are predicted to be detected by LISA with high signal-to-noise (SNR) and to aid their gravitational-wave (GW) parameter estimation. 
On the other hand, as a class of binary with the shortest orbital period, UCBs represent the terminal phase of some binary evolution, which provide opportunities in studying physics under extreme conditions and give crucial constraints on the binary evolution, such as mass-accretion/loss processes, common-envelope evolution, and angular-momentum loss mechanisms (see also \citealt{Toonen+etal+2014,Wang+etal+2021,Chen+etal+2020,Zhu+etal+2012,Rappaport+etal+1983}).

Noninteracting black hole binaries (or candidates), which cannot be detected by current X-ray detectors, have been discovered by periodic photometric variability and radial velocities (RVs) of their visible companion stars \citep{Thompson+etal+2019+Science,Liu+etal+2019+Nature}. 
Furthermore, some newest researches suggest that the black holes in the short-period ellipsoidal variables can be revealed by analyzing the Fourier amplitudes of their light curves \citep{Gomel+etal+2021+ellipsoidals_I,Gomel+etal+2021+ellipsoidals_II,Gomel+etal+2021+ellipsoidals_III}. 
As the cross field between GW verification binaries and black hole binaries, the ultracompact black hole (X-ray) binaries \citep{Bahramian+etal+2017+UCBHXB}, in which the X-ray radiation should be inefficient \citep{Knevitt+etal+2014,Menou+etal+1999}, are expected to be first discovered by high-cadence, wide-area optical survey missions or next-generation gravitational-wave observatories. 
The ultracompact black hole binaries could be the unique Galactic black hole systems that can be detected by both gravitational and electromagnetic waves, implying they will be the most direct evidence that the stellar black hole exists.

\begin{figure*}
    \includegraphics[width=0.94\textwidth]{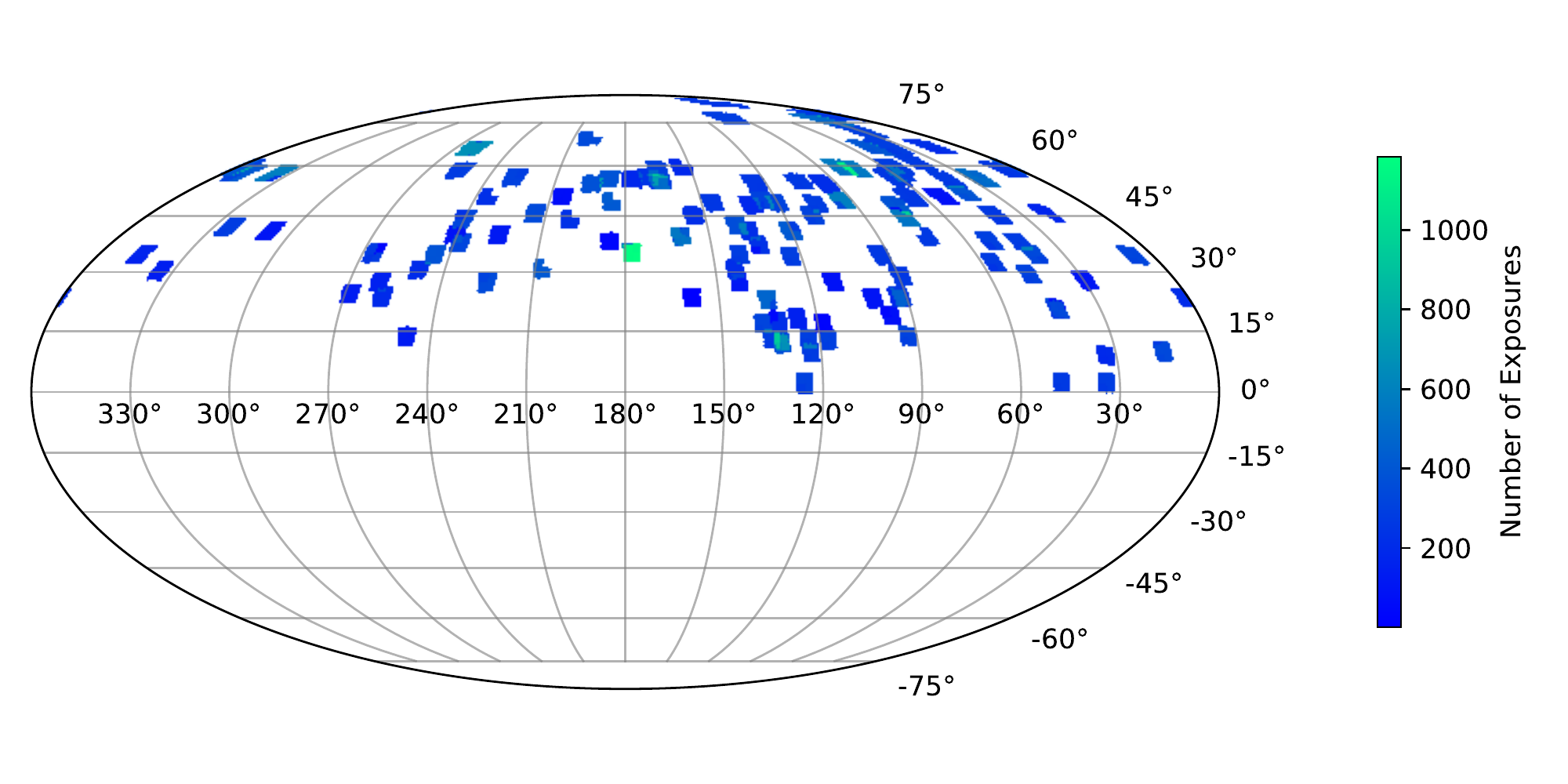}
    \caption{
     \textbf{Observation sky areas of the TMTS shown in equatorial coordinates.}
     The sky map is plotted by using the HEALPix package (\url{http://healpix.sourceforge.net}) with NSIDE=128 \citep{Gorski+etal+2005}.
     The depth of the color represents the total number of 1-min exposure.
     } 
    \label{fig:skymap}
\end{figure*}

So far, a dozen of ground-based survey missions have operated to search for transients and variables on different timescales. 
These missions include the Deep Lens Survey (DLS, 1999--2005, \citealt{DLS+2004_short}), the Faint Sky Variability Survey (FSVS, \citealt{FSVS+2003+reduction})
the RApid Temporal Survey (RATS, \citealt{RATS+2005+survey} ), 
the Catalina Real Time Survey (CRTS, since 2007, \citealt{Catalina+2009+first,Catalina+2014+ultracompact,Catelina+2014+CV} ), the Palomar Transient Factory (PTF, 2009--2012, \citealt{PTF+2009+performance,PTF+2009+OT}), the omegaWhite survey ( \citealt{OmegaWhite+2015+survey} ), the Intermediate Palomar Transient Factory (iPTF, 2013--2017, \citealt{iPTF+2019+detectability,Ho+etal+2018+iptf}), 
the High Cadence Transient Survey (HiTS, \citealt{HiTS+2018}), 
the Evryscope (since 2015, \citealt{Evryscope+2019+performance}), the Zwicky Transient Facility (ZTF, since 2017, \citealt{ZTF+2019+first,ZTF+2019+products}), the Compact binary HIgh CAdence Survey (CHiCaS, \citealt{CHiCaS+2020+survey}).
High-cadence surveys, especially uninterrupted time series photometry, are more efficient in discovering short-period light variations and phenomenon associated with stellar flares/bursts. 
However, only a few missions (e.g. ZTF high-cadence Galactic Plane Survey, \citealt{Kupfer+etal+2021}) insist on performing uninterrupted photometry, as high-cadence surveys will significantly sacrifice the coverage of the sky area. 
Fortunately, several space-based survey missions, such as the \emph{Kepler} mission \citep{Kepler+2010+first,Koch+etal+2010+Kepler_intrument} and the Transiting Exoplanet Survey Satellite (TESS,\citealt{Ricker+etal+2014+TESS,TESS+2015}), have operated to perform long-duration uninterrupted photometry. But \emph{Kepler} was limited to observe only Cygnus-Lyra region and ecliptic plane, while TESS was designed to monitor the brightest dwarf stars. Moreover, these space-based missions usually suffer low efficiency of data transportation, and thus finally provide light curves for only hundred thousands of objects.  

As \cite{Burdge+etal+2020+systematic} mentioned, a systematic search and study of ultracompact binaries relies not only on the high-cadence photometry, but also on the time-resolved spectroscopy. The high-cadence photometry is used to search for periodic signals, while the spectra are used to determine semi-amplitude of radial velocities.  On the other hand, studies of fast-evolving transients such as flare stars also requires spectroscopic 
confirmation \citep{Kulkarni+FOTs+DLS+2006,Ho+etal+2018+iptf}.
Therefore, we initiated a new high-cadence survey mission, with an attempt to cover the LAMOST sky areas 
with the Tsinghua University-Ma Huateng Telescopes for Survey (TMTS, \citealt{TMTS+2020+survey}). The LAMOST has started
the time-domain medium-resolution spectroscopic survey since October 2018 \citep{Liu+etal+2020}, which provides precise measurements of the RV variations for stars brighter than 15.0 mag.

In this paper, we present the methods of data analysis and preliminary results for the first-year high-cadence surveys from the TMTS. 
The schema of first-year observations and light-curve dataset are described in Section~\ref{sec:observation}. 
The descriptions of photometry and calibration are presented in Section~\ref{sec:reduction}.
In Section~\ref{sec:methods}, we introduce the methodology of detecting variability, periodicity and flares in the TMTS light curves, respectively. In this section, we also describe the source selection with the Hertzsprung–Russell (HR) diagram. 
We present some selected results in Section~\ref{results}.

\section{Observation}
\label{sec:observation}

TMTS is a multiple-tube telescope system consisting of four {40-cm} optical telescopes with a total field of view (FoV) of about 18~deg$^2$ { (4.5~deg$^2$ for each telescope ) and a plate scale of $1.86^{\prime \prime}\,$pixel$^{-1}$.}
The TMTS system is equipped with 4096$\times$4096 pixels CMOS cameras, which have short read-out time ($< 1$~s) and allow to conduct high-cadence photometry for targets on large sky areas. {Detailed introduction about the performance of TMTS is described in \cite{TMTS+2020+survey}.}

Since TMTS and LAMOST have similar FoV and locate at the same site (i.e. Xinglong Station of NAOC), the former is an ideal telescope system to carry out collaborative tasks with the latter. 
{
At Xinglong Station, the typical seeing is better or comparable to $2.6^{\prime \prime}$ for 80\% of nights and the sky brightness at zenith is around 21.1~mag/arcsec$^2$ \citep{Huang+etal+2012+TNT,Zhang+etal+2015+Xinglong_conditions}. Due to the light pollution from the surrounding cities, the sky brightness increases with the increase of zenith angle.
As \cite{Zhang+etal+2015+Xinglong_conditions} introduced, 32\% of nights in Xinglong Station have cloud-free observations for at least 6 hours, which means that about 117~nights per year are suitable for the interrupted photometric observations required by the TMTS.}

The TMTS has two observation modes:  (i) staring at the LAMOST areas for the whole night whenever possible with a cadence of about 1 minute; (ii) supernova survey with a cadence of about 1--2 days. In this paper, we concentrate on the first-year observations of the LAMOST sky areas. 
In order to achieve a high signal-to-noise ratio, the observations are conducted in Luminous filter {(L filter hereafter), which has a very wide coverage ranging from 330~nm to about 900~nm when combined with the CMOS detector (see Figure~6  in \citealt{TMTS+2020+survey}). 
Similar to \emph{Gaia}'s G band (330--1050~nm, \citealt{Gaia_collaboration+2018+data}), the ``white-light'' band can maximize the detection depth of optical telescopes. For a 1-min exposure, the 3-$\sigma$ detection limit of the TMTS can reach about 19.4~mag.}

\begin{figure*}
    \includegraphics[width=0.49\textwidth]{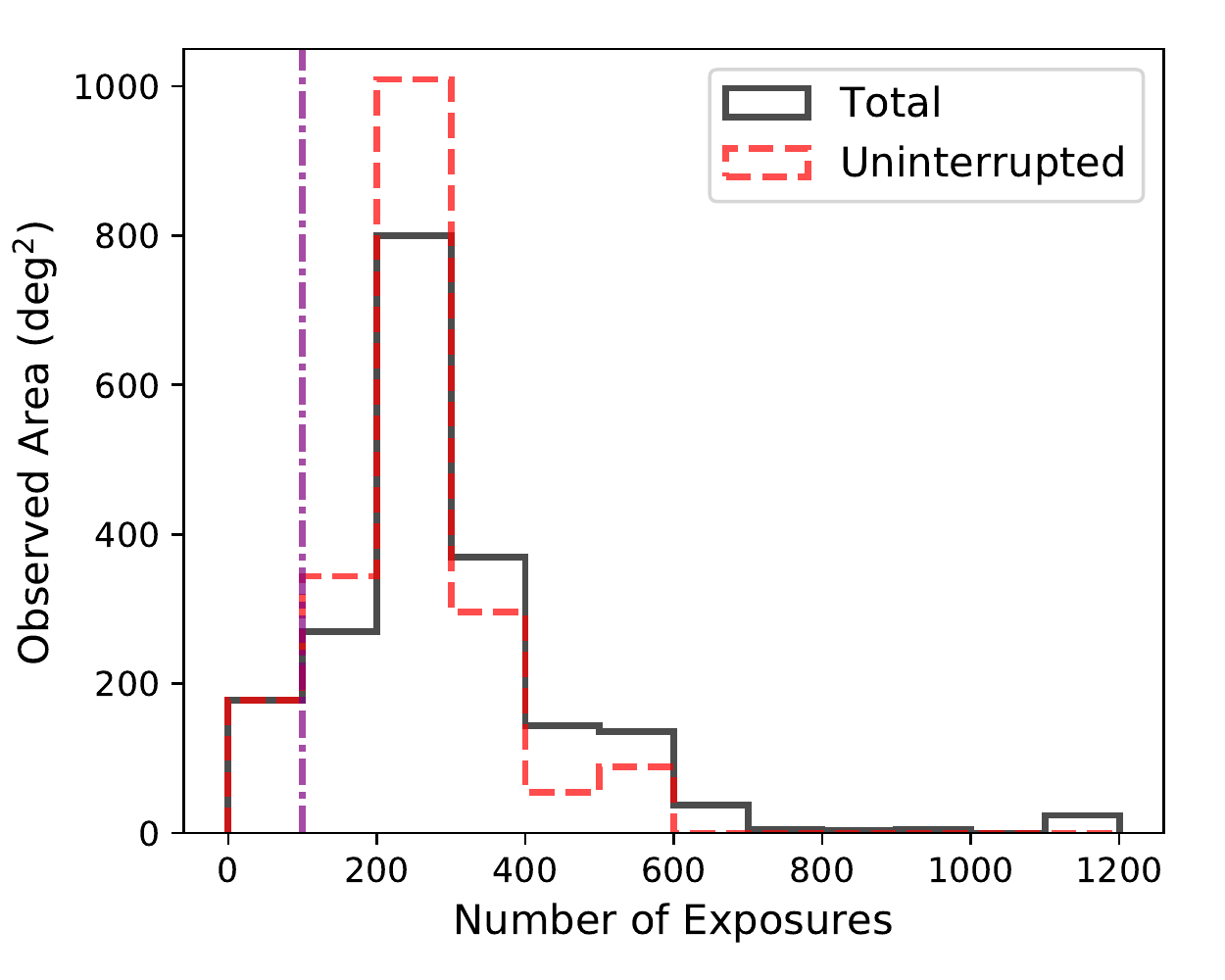}
    \includegraphics[width=0.49\textwidth]{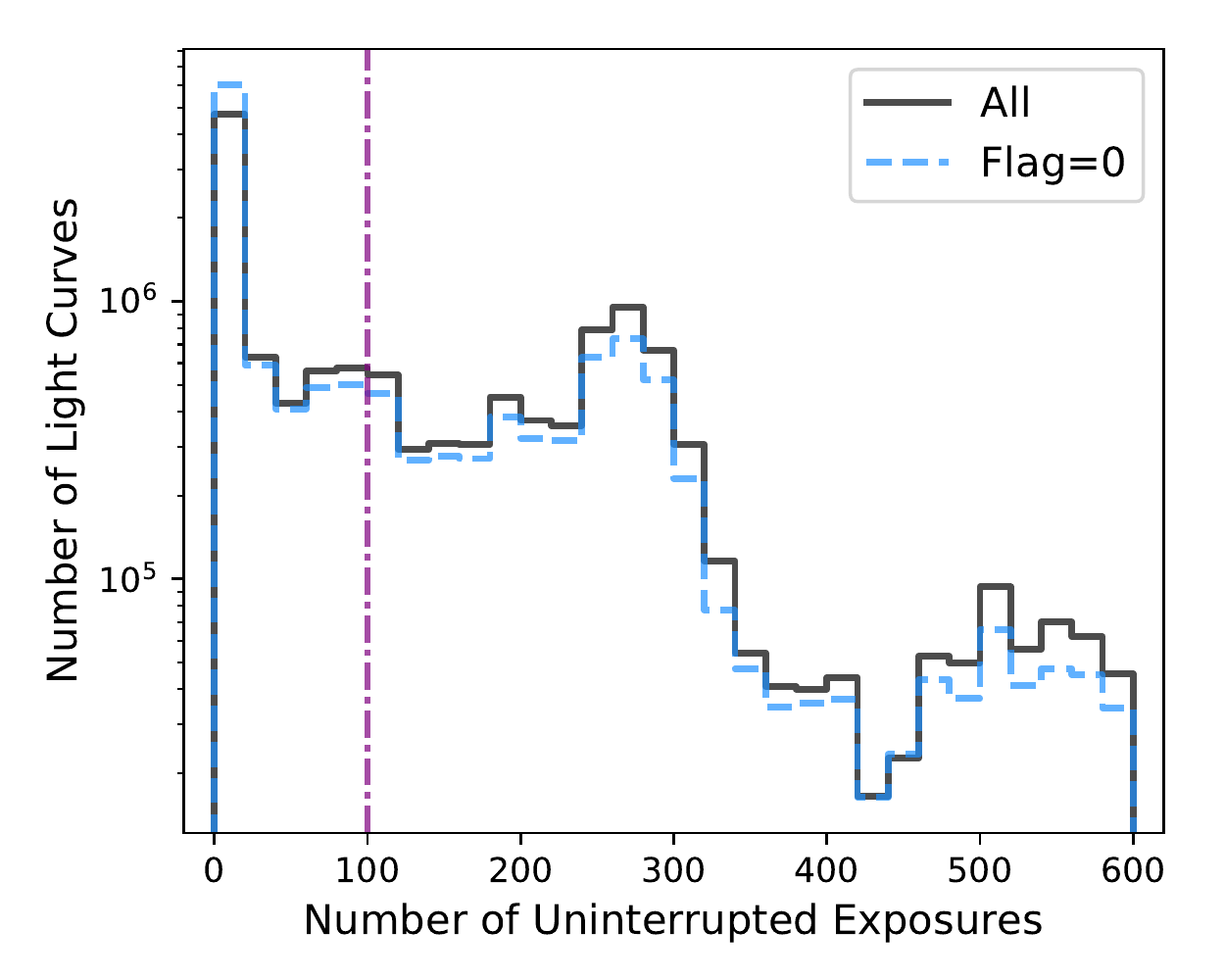}
    \caption{
     \textbf{Histogram of number of (1-min) exposures for observed area (left) and light curves (right).}  \emph{Left:} The black solid line and red dashed line represent the statistics based on overall and uninterrupted observations, respectively. The purple dot-dashed lines indicate the cut-off value (i.e. 100 repeated exposures). \emph{Right:}  The blue dashed line represents the exposure number of valid measurements (i.e. flag$=0$, see details in Section~\ref{sec:reduction}). 
     } 
    \label{fig:obs_stat}
\end{figure*}

As shown in Figure~\ref{fig:skymap}, the TMTS observed 188 LAMOST plates during the whole year of 2020, covering a total sky area of $\approx 1970~{\rm deg}^2$.
Among them, the sky area of $\approx1793$ deg$^2$ has at least 100 uninterrupted 1-min exposures, as shown in the left panel of Figure~\ref{fig:obs_stat}. 
Notice that, the 1-min image here is combined from six 10-second images and its frame rate thus dropped to $\approx 1/75$~Hz. For the purpose of selecting variables based on the light-curve analysis, we focus on the observed sky areas with at least 100 epochs, which takes up about 96\% of the LAMOST sky areas monitored during the first year. 
The high-cadence survey allows us to discover and identify variables in the LAMOST fields on a time scale of about 1 minute.

As can be seen from the right panel of Figure~\ref{fig:obs_stat}, the TMTS has produced $\approx$13 million uninterrupted light curves during the survey conducted in 2020, of which {$\approx$6 million} have at least 100 repeated measurements. Notice that there are about 4.7 million light curves with less than 20 epochs. The sources with such sparse measurements could locate either near the edge of FoV or hover around the detection limit.
Based on the light curves with at least 100 ``valid'' measurements (see details in Section~\ref{sec:reduction}), we built a dataset including 4.9 million selected light curves from the first-year survey,  namely \textbf{\tmtsdata} dataset.
It is worth noting that, multiple light curves may correspond to the same source due to that some sources are located in the overlapping FoV of multiple telescopes of TMTS.
Since each light curve can be used to detect variability and periodicity  for a source independently and only a small part of sources have multi-epoch light curves, the repeated light curves are not spliced together.

\section{Photometry and Calibration}
\label{sec:reduction}

All of the 10-sec raw images from the TMTS are first bias-, dark- and flat-corrected using the \emph{FITSH} package \citep{Pal+2012+FITSH}.
Then the astrometric calibration is applied to the 10-sec single frame using the soft package \emph{SCAMP} \citep{Bertin+2006+SCAMP} and the reference catalog of PPM-Extended (PPMX, \citealt{Roser+etal+2008+PPMX} ). The \emph{SCAMP} can automatically generate accurate World Coordinate System (WCS) information by cross-correlating the reference catalog, and it can give accurate astrometric solutions for the FITS images.
To improve the detection depth, 6 successive single frames are median combined into a 1-min image using the soft module \emph{SWarp} in the \emph{TERAPIX} pipeline \citep{Bertin+etal+2002+TERAPIX}.
We extract the fluxes of sources on the combined images using the soft package Source Extractor (SExtractor, \citealt{Bertin+Arnouts+1996+SExtractor}).

We checked the SExtractor flag for all of the TMTS measurements. The SExtractor flag $\ne 0$ means that there are some problems in the measurements, e.g. blending or saturation (see details in \url{https://sextractor.readthedocs.io/en/latest/Flagging.html}). 
In addition, we added a new flag bit (value$=256$) to mark the measurements within 100 pixels of the detector boundary,
as these measurements frequently cause spurious variations in the light curves and are difficult to be calibrated.
Due to immature manufacturing process, the backgrounds of four regions divided by the X/Y midlines in the CMOS detector are not completely consistent, especially during big moon nights (see \citealt{TMTS+2020+survey}).This inconsistency would cause spurious variation in the light curves of objects across the midlines, we thus added an additional flag bit (value$=512$) to those detections within 40 pixels of the detector midlines. 
The histogram of flag=0 measurements (``valid measurements'' hereafter) is also shown in the right panel of Figure~\ref{fig:obs_stat}, and the number of light curves with at least 100 repeated valid measurements is $\approx4.9$ million.

\begin{figure}
    \includegraphics[width=0.47\textwidth]{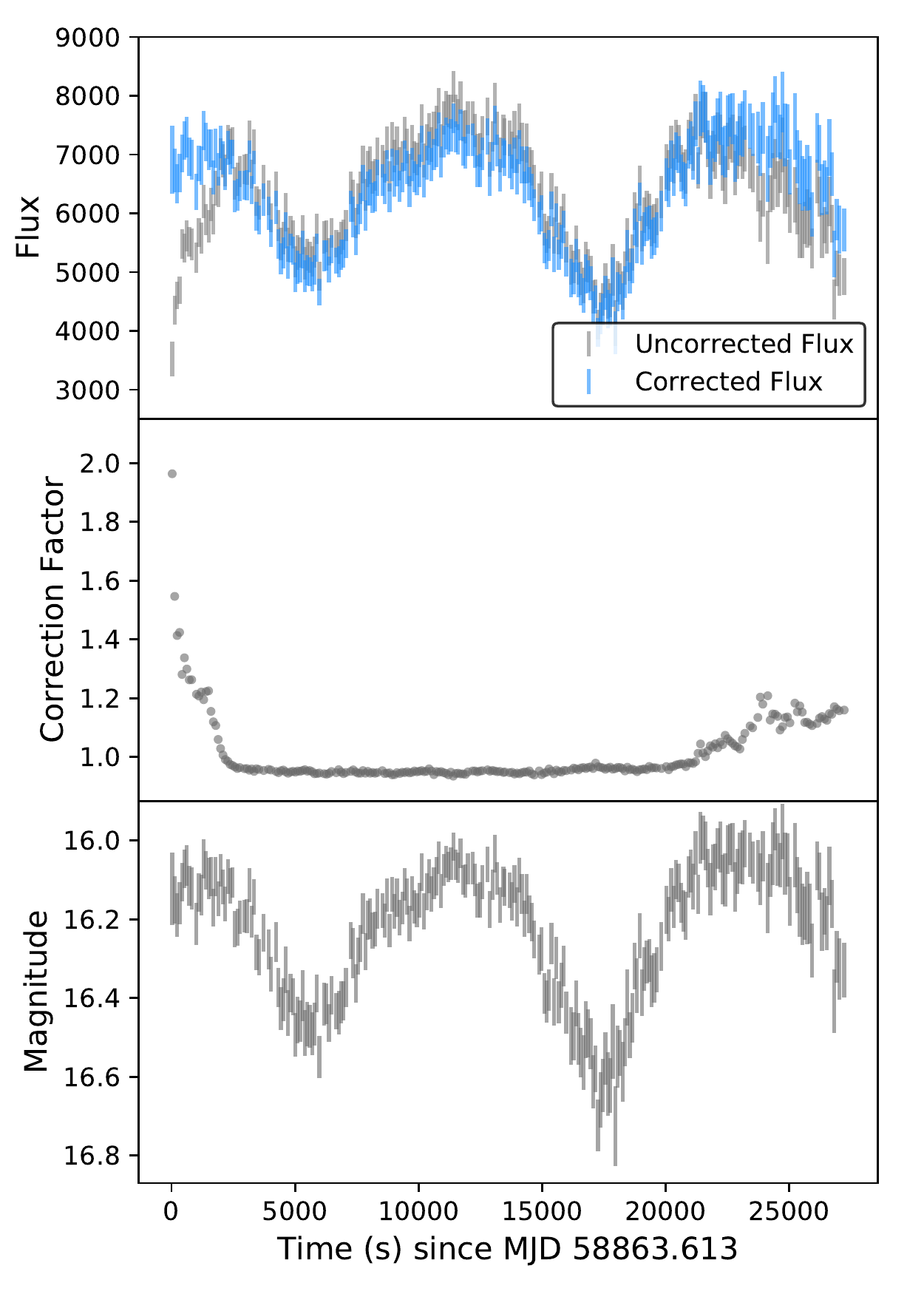}
    \caption{
     Example of TMTS light curve and correction factor (i.e. the $\alpha_{i}$ in Eq~\ref{Eq+corrected_flux}) for the W UMa-type eclipsing binary CRTS J075625.0+420405 \citep{Drake+etal+2014+catalina_period,Marsh+etal+2017+eb}, which was observed by the telescope \#3 of TMTS for $\approx$5.3 hours on January 15, 2020.} 
    \label{fig:lc_example}
\end{figure}

The flux measurements from continuous observations 
were combined into a light curve. Before detecting real variability and periodicity, we need to first remove the systematic effects like spurious variations caused by changes in airmass, lunar phase and solar altitude etc. 
The \emph{Tamuz}'s method \citep{Tamuz+etal+2005+sysremove} and the Principal Component Analysis (PCA)  are not adopted in the analysis. Because the extinction coefficients in these algorithms cannot be correctly determined for variable stars, as these methods actually make an assumption that the magnitudes of each light curve are equal to its average. Some methods can better constrain the coefficients for variable stars but a prior model on the intrinsic variation needs to be assumed (see also \citealt{Aigrain+etal+2017+sysrem,Ofir+etal+2010+corot}). 
For these reasons, we developed a weighted version of ``differential photometry'' to reduce the systematic errors of the light curves.  Similar to the \emph{Tamuz}'s method and PCA, our algorithm adopted in the analysis also involves removing the common trends and features among a large set of light curves. These common trends and features can be modeled by (weighted) averaging all measurements for constant stars (i.e., the stars that show no variations during the observations) within the FoV. The ``effective extinction coefficient'' for each star (i.e $c_i$ in \citealt{Tamuz+etal+2005+sysremove}) is not set in our algorithm, since the coefficient cannot be determined accurately for those variable stars.
Instead, a weighted factor based on the separations between reference stars of constant luminosity and target object is introduced to modulate various effects in different detection regions. Hence, the corrected flux $F_{i}^{\rm corr}$ at epoch $t_i$ for target source $i$ is calculated as
\begin{equation}
 F_{i}^{\rm corr}=F_i\times \alpha_i= F_i\times \prod\limits_{j=1}^M (\frac{\overline{F_{j}^{\rm ref} }}{   F_{i,j}^{\rm ref} })^{\omega_j/\sum\limits_{j=1}^M \omega_j  }
 \label{Eq+corrected_flux}
\end{equation}
where $\alpha_i$ represents the correction factor and $M$ is the total number of reference stars.
$F_i$ and $F_{i,j}^{\rm ref}$ are the uncorrected flux for the target and the $j$th reference star at time $t_i$, respectively.  $\overline{F_{j}^{\rm ref} }$ is the average uncorrected flux of the $j$th reference star over an observation night, and it is expressed as $ \sum\limits_{i=1}^N  F_{i,j}^{\rm ref}/N$, where $N$ is the number of epochs. The $\omega_j$ is a weighted factor for the $j$th reference star, which is set to $1~{\rm arcsec^2}/(a_j+C)^2$, where $a_j$ is the separation between the target and the $j$th reference star. 
The characteristic separation $C$ is set to be a small value ( i.e. $60$~arcsecs in our work) to avoid the singular value when the reference star is very close to the target.
The corrected fluxes seem to be insensitive to the value of $C$
and the results are not significantly different even if the characteristic separation $C$ is set as 10~arcmins.

The sources with 13 mag < $G$ < 17 mag were selected to be the reference star candidates, the $G$ represents the mean G-band magnitudes from Gaia DR2 database \citep{Gaia_Collaboration+2016+performance,Gaia_collaboration+2018+data}.
In order to improve the calibration process, we only adopted the reference star candidates with $q=100\%$ for their light curves, where $q=N({\rm flag=0})/N$ is a parameter to evaluate the quality of a light curve. $N$ indicates the total number of epochs for a given light curve and $N({\rm flag=0})$ represents the number of valid measurements.
Notice that, the reference stars may contain some variables which should be revealed and kicked out through iterative process.
In order to reveal those variables in the reference stars, we calculated the inverse von Neumann ratio for all of our light curves (see \citealt{Shin+etal+2009+variable,Sokolovsky+etal+2017+variables}). This ratio is a very useful variability index derived by testing the independence of successive measurements. The inverse von Neumann ratio \citep{Sokolovsky+etal+2017+variables}  is defined as 
\begin{equation}
\frac{1}{\eta}=\frac{\sum\limits_{i=1}^{N} (F_{i}^{\rm corr}-\overline{F^{\rm corr}} )^2}
{\sum\limits_{i=1}^{N-1}  (F_{i+1}^{\rm corr}-F_{i}^{\rm corr})^2 }
\label{Eq+neumann_ratio}
\end{equation}
where $\overline{F^{\rm corr}}$ represents the corrected flux averaged over all epochs. We set a very tight cut-off value, i.e. 0.8, to exclude all variables from the reference stars. We will explain why $\frac{1}{\eta}=0.8$ is a robust threshold in Section~\ref{sec+variable_detection}.

\begin{figure}
    \includegraphics[width=0.47\textwidth]{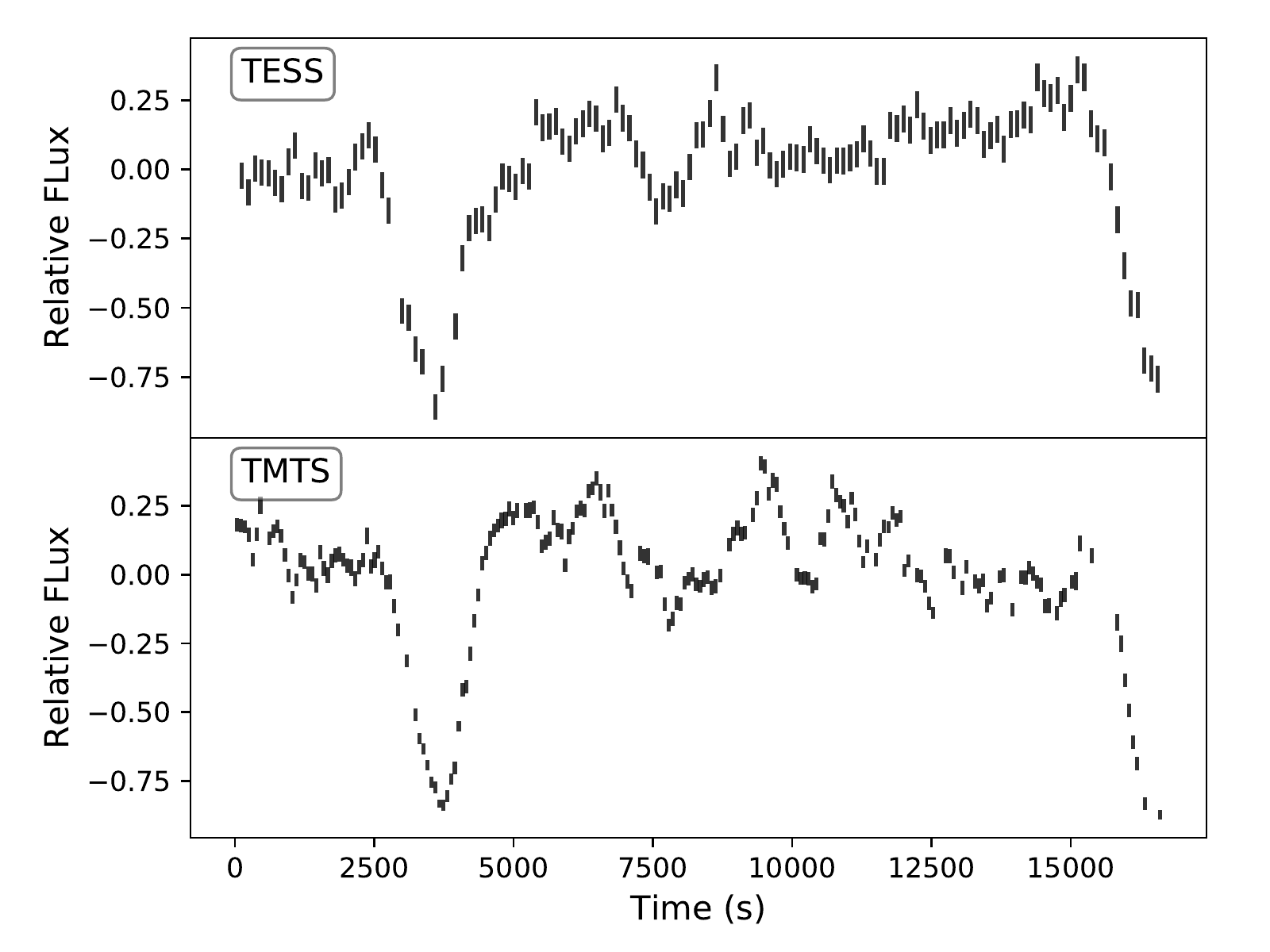}
    \caption{
   \textbf{A comparison between TESS and TMTS light curves.}
   The observed object is HS 0455+8315, which is an eclipsing cataclysmic variable with a visual magnitude from about 15 to 17 \citep{Downes+etal+2001+CV}.
   TMTS observed this object on November 2, 2020, and the TESS (PDC) light curve was obtained from the observations on June 9, 2020 (Sector~26).
   The start time of TESS light curve is reset to make its major eclipse coincide with that of TMTS.
    } 
    \label{fig:tess_tmts_lc}
\end{figure}

An example of re-calibrated flux and correction factor for the TMTS light curve of a WUMa-type eclipsing binary (i.e. CRTS J075625.0+420405) are shown in the top and middle panels of Figure~\ref{fig:lc_example}, respectively. To obtain the corresponding magnitudes, we also calculate the magnitude zero point $m_0$ for each targets, which follows as
\begin{equation}
m_0=\frac{\sum\limits_{j=1}^M \omega_j\times (2.5\, \log_{10} \overline{F_{j}^{\rm ref}}  + G_j)}{\sum\limits_{j=1}^M \omega_j}
\label{Eq+zero_magnitude}
\end{equation}
where $G_j$ is the Gaia DR2 G magnitude of the $j$th reference star. The magnitude obtained at epoch $t_i$ is thus estimated as $ m_i=-2.5\times \log_{10}F_{i}^{\rm corr} + m_0$. Inserting Equations~\ref{Eq+corrected_flux} and \ref{Eq+zero_magnitude} into the above equation, the magnitude can be expressed as 
\begin{equation}
m_i=\frac{\sum\limits_{j=1}^M \omega_j\times (-2.5\, \log_{10}\frac{F_i}{ F_{i,j}^{\rm ref}}+ G_j)}{\sum\limits_{j=1}^M \omega_j}~.
\end{equation}
The bottom panel of Figure~\ref{fig:lc_example} shows the final magnitudes obtained with the TMTS for CRTS~J075625.0+420405.
It is worth noting that, the measurement accuracy of TMTS is superior to the space-based survey mission TESS, as shown by the comparison of the light curves obtained for the same source (see Figure~\ref{fig:tess_tmts_lc}).

\begin{figure}
    \includegraphics[width=0.47\textwidth]{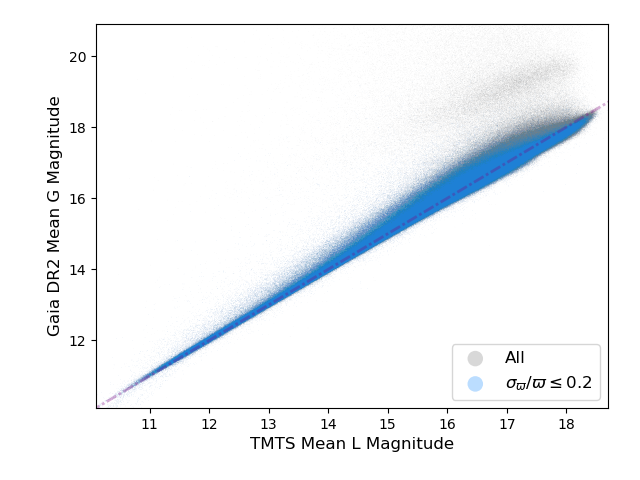}
    \caption{
   \textbf{Comparison of TMTS L-band magnitudes with Gaia DR2 G-band Magnitude.}
   The blue points represent the sources with reliable parallax measurements ( $\sigma_\varpi/\varpi \leq 0.2$), which means that these sources have more precise astrometric solutions. The purple dot-dashed line represents the diagonal line.
    } 
    \label{fig:gaia_tmts_mag}
\end{figure}

Figure~\ref{fig:gaia_tmts_mag} shows the comparison of TMTS magnitudes obtained in L band with the G magnitudes from Gaia DR2. The mean TMTS magnitudes here were taken from the light curves of \tmtsdata.
The corresponding Gaia sources with reliable parallax measurements ($\sigma_\varpi/\varpi \leq 0.2$ here, where $\varpi$ is the parallax and $\sigma_\varpi$ represents the error of parallax ) are used to cross-match the TMTS sources.
One can see that TMTS L magnitudes are basically consistent with the Gaia G magnitudes while the scatter between these two magnitude systems tends to increase at the faint end (see the blue points in the Figure~\ref{fig:gaia_tmts_mag}). 
Notice that the Gaia sources with spurious parallax values (e.g. negative value) are usually faint and likely locate in crowded regions (e.g. low Galactic latitude) where their astrometric solutions are poorly constrained \citep{Gaia_collaboration+2018+data}.
For the comparison, we also showed an overall version that includes the sources with spurious astrometric measurements ( see the grey points in the Figure~\ref{fig:gaia_tmts_mag}), these sources with poor astrometry from Gaia caused an additional cluster above the original distribution when matching with the TMTS sources, which also appeared in the comparison between Gaia DR1 G magnitudes and DR2 G magnitudes (see details in \url{https://gea.esac.esa.int/archive/documentation/GDR2/index.html}).

\section{Methods}
\label{sec:methods}
\subsection{Variability detection}
\label{sec:variability_detection}

\begin{figure}
    \includegraphics[width=0.47\textwidth]{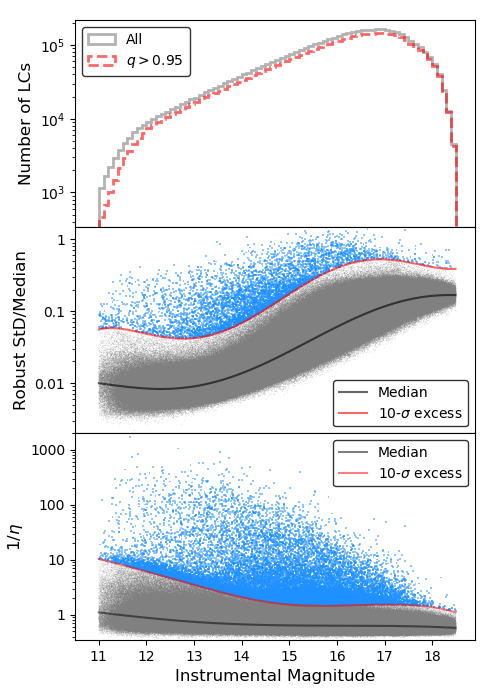}
    \caption{
    \textbf{Distribution of light curves, robust Std and inverse von Neumann ratio versus the instrumental magnitude.}
     \emph{Upper panel}: Distribution of the number of the TMTS light curves from the \tmtsdata dataset against the magnitudes. The red dashed line represents the light curves with quality higher than 95\% { (see Section~\ref{sec:reduction}) }. The bin size of the histogram is 0.1 mag.
     \emph{Middle panel}: The normalized robust StD versus the magnitude.
    The black and red solid lines represent the polynomial fit to the median and 10-$\sigma$ threshold, respectively.
      The blue squares indicate those light curves with variability index being above the 10-$\sigma$ threshold.
     \emph{Lower panel}: The inverse von Neumann ratio versus the magnitude.
     } 
    \label{fig:variability_detection}
\end{figure}
\label{sec+variable_detection}

Difference image analysis (DIA, \citealt{Tomaney+Crotts+1996+DIA,Alard+Lupton+1998+DIA}) and light curve analysis (LCA, \citealt{Sokolovsky+etal+2017+variables}) are two main methods to search for variables. Compared with the DIA, the light curve analysis, based on the measurements obtained at more than two epochs, can reveal low-amplitude variability.
In search for variable sources from the TMTS light curves, we calculated two common variability indices. 
As it is difficult to detect reliable variabilities for the light curves covering a very short duration, we thus identified variable sources for those light curves with at least 100 valid epochs (see also \citealt{Gomel+etal+2021+ellipsoidals_III,Kupfer+etal+2021}).

We selected the TMTS light curves of the \tmtsdata dataset ($\approx$4.9 million)  with instrumental magnitude $ 11.0 < \widetilde{m} <18.5 $. 
Since the average zero point of all measurements is $25.59\pm0.25$, we defined the instrumental magnitude as $\widetilde{m}=-2.5\times\log_{10}( \overline{F^{\rm corr}} )+25.6$.
{
The value of the instrumental magnitude here is close to but not equal to the astrophysical magnitude, due to that the variations of photometric zero points with sky areas and observation conditions are not considered.}

The upper panel of Figure~\ref{fig:variability_detection} shows the histogram of number of TMTS {light curves} from the \tmtsdata dataset as a function of the instrumental magnitude. From magnitude 12 to 17, the number density of TMTS {light curves} increase by about an order of magnitude. The highest number density appears at $\widetilde{m} \approx 17.0$~mag. At the fainter end, the decrease in number density is primarily due to that the detection depth varies with observation conditions; at the brighter end, the detections suffer from effects of both saturation and small number of bright stars. 

We have calculated the robust standard deviation (StD) and inverse von Neumann ratio ($1/\eta$ here) as a function of their instrumental magnitude $\widetilde{m}$ for the selected TMTS light curves (see the middle and lower panels of Figure~\ref{fig:variability_detection}). The robust StD is the standard deviation inferred from the central 50 percentile of the data points by assuming a Gaussian distribution \citep{ZTF+DR1+2020}, implying that the robust StD is extremely insensitive to outliers or occasional variations. The normalized robust StD, defined as the ratio of the robust StD to the median flux, increases from $\sim 0.01$ to $\sim 0.1$ when the brightness of the sources decreases from 12 mag to 18 mag.
Those points that have significantly larger ``scatter'' than the expected are very likely due to light variations. We use 5th-order polynomial to fit the median and the 10-$\sigma$ excess, respectively, which are both calculated in a bin of 0.1 mag. 
There are about 5,600 light curves have higher robust Std than the 10-$\sigma$ threshold (the blue square in the middle panel of Figure~\ref{fig:variability_detection}).
These light curves correspond to about 5,300 GAIA DR2 sources. However, it is difficult to conclude that these sources are all astrophysically variable stars, since blended sources can also show variabilities in their light curves.
For example, \cite{Kupfer+etal+2021} recently revealed a false positive rate ({i.e. the rate of non-astrophysically variable sources) of up to 85\%} for the variability detection in the high-cadence Galactic Plane observations of ZTF.
{
By visually inspecting 300 TMTS light curves with significance of light variations being above the 10-$\sigma$ threshold, we found that non-astrophysically variable stars accounted for about 67\%. For a lower threshold, i.e. 5$\sigma$, about 23,000 light curves can be selected but the false positive rate increases to about 86\%. Obviously, the lower thresholds can be used to pick more astrophysically variable stars, but the higher false positive rates also result in a huge sample containing more non-astrophysically variable sources which is hard to be visually inspected.
Therefore, we will take the variability indices as an auxiliary condition to select periodic variables and flare stars.
}

\begin{figure}
    \includegraphics[width=0.47\textwidth]{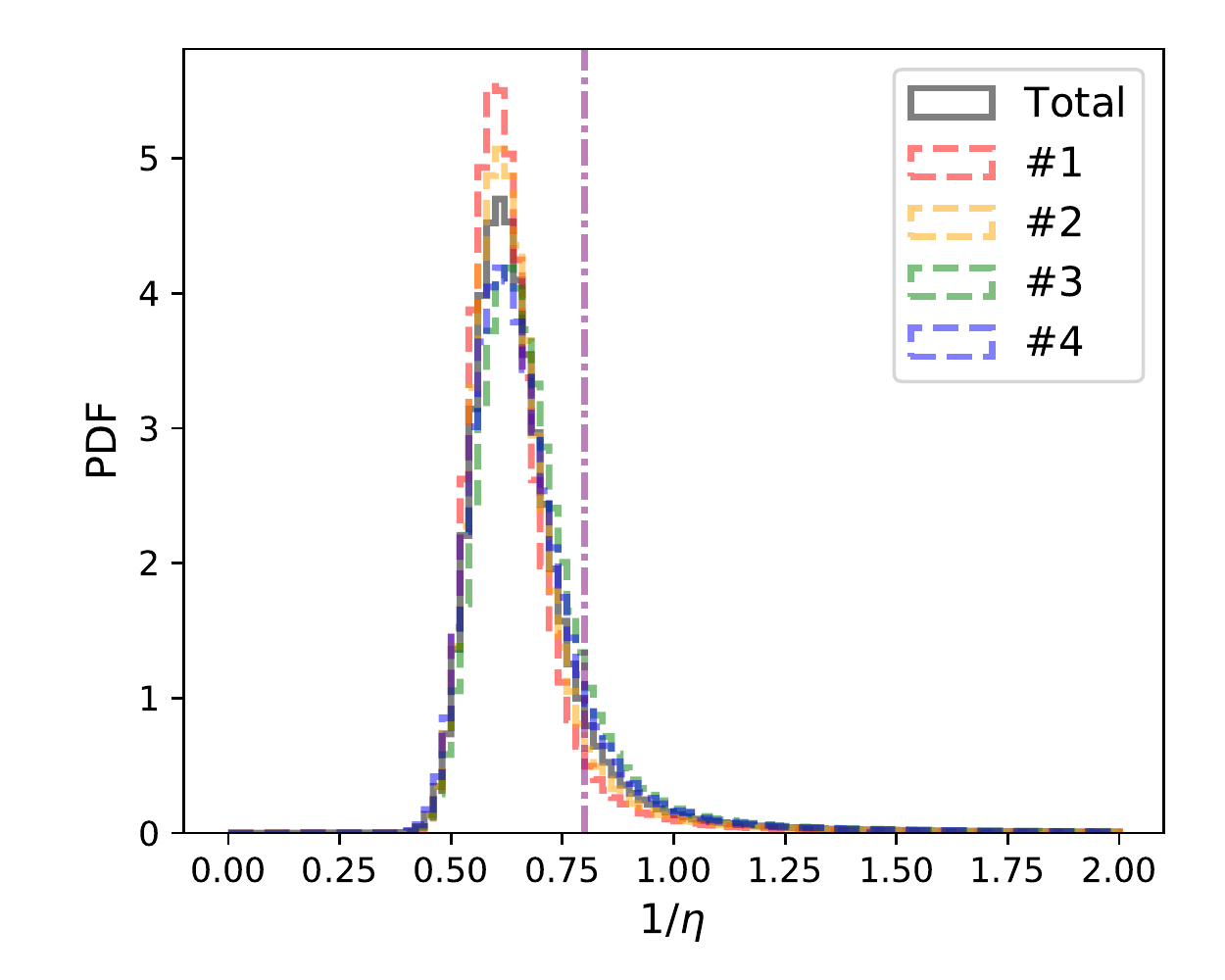}
    \caption{
    \textbf{Density distribution of inverse von Neumann Ratio $1/\eta$ of light curves from the \tmtsdata dataset for the all four (grey solid line) and each of four (colorful dashed line) telescopes of TMTS.}
    The bin size here is 0.02.
    The purple dot-dashed line indicate the cut-off value to determine the non-variable reference sources for photometric calibration. 
     } 
    \label{fig:neumann_dist}
\end{figure}

The inverse von Neumann ratio quantifies the smoothness of a time-series successive variation and does not depend on the uncertainty of the measurements as the contribution of uncertainty has been nearly {offset} by its denominator as shown in Eq.~\ref{Eq+neumann_ratio}. For an ideal time series of photometry following a Gaussian distribution, the expected value of its $1/\eta$ is equal to 0.5 .
However, for real photometric measurements, which do not follow the Gaussian distribution or are not completely independent of each other, the cut-off value should be determined based on the distribution of $1/\eta$ (see \citealt{Sokolovsky+etal+2017+variables}). 
Given the characteristics of $1/\eta$, we use a 3rd-order polynomial (rather than the 5th-order polynomial) to fit the median and robust StD versus $\widetilde{m}$, respectively,  which yields $\approx$24 thousand light curves showing variations beyond the 10-$\sigma$ threshold. 
Notice that the $1/\eta$ does not obey the Gaussian distribution in practise, hence the ``$\sigma$'' here does not represent the confidence level corresponding to Gaussian distribution.
Moreover, we found that $1/\eta$ tends to be larger for brighter sources, which could be caused by the unmarked saturations. The saturation effect tends to reduce the independence of successive measurements and thus increase the value of $1/\eta$.

For the purpose of excluding all variable stars from the reference stars, we use $1/\eta$ to identify variable sources because the robust StD parameter is insensitive to the occasional variations (e.g. the variations of flare stars).
Since the threshold of variability index varies with the instrumental magnitude, we thus defined a statistic parameter $\epsilon_{\frac{1}{\eta}}=[\frac{1}{\eta}- \nu(\widetilde{m}) ]/\sigma (\widetilde{m})$, where $\nu(\widetilde{m})$ and $\sigma (\widetilde{m})$ are the median and robust standard deviation of inverse von Neumann ratios for the TMTS light curves at a given magnitude, respectively.
The $\epsilon_{\frac{1}{\eta}}$ is a key parameter to introduce the significance of variability for a light curve.

As introduced above, although the identifications of astrophysically variable stars using the the variability indices has a low true positive rate (TPR), they can be used to identify stars of constant luminosity at a very high TPR since the light variations caused by image quality will not increase the FPR of non-variable sources. 
The setting of the thresholds for variability indices is usually arbitrary (see \citealt{ZTF+2019+first,Nidever+etal+2021+survey,Kupfer+etal+2021}).
In the photometric calibration process (shown in Section~\ref{sec:reduction}), 
a fixed threshold $1/\eta\leq0.8$ is empirically set to identify non-variable sources.
It is a tight threshold that ensures $\epsilon_{\frac{1}{\eta}} \lesssim 1.5$ for the reference stars in all observed magnitudes, corresponding to the exclusion of about 12\% of the reference star candidates.

Furthermore, we compare the density distribution of inverse von Neumann ratio for each telescope of the TMTS system. 
As Figure~\ref{fig:neumann_dist} shows, the $1/\eta$ distribution of each telescope is roughly consistent with each other except that the telescope \#1 and telescope \#2 have a slightly more concentrated distribution, implying that the capability of variability detection is almost equivalent for each telescope. Therefore, the same threshold for variability indices is adopted for all of the four telescopes.

\subsection{Periodicity detection}
\label{sec:periodicity_detection}

\begin{figure}
    \includegraphics[width=0.47\textwidth]{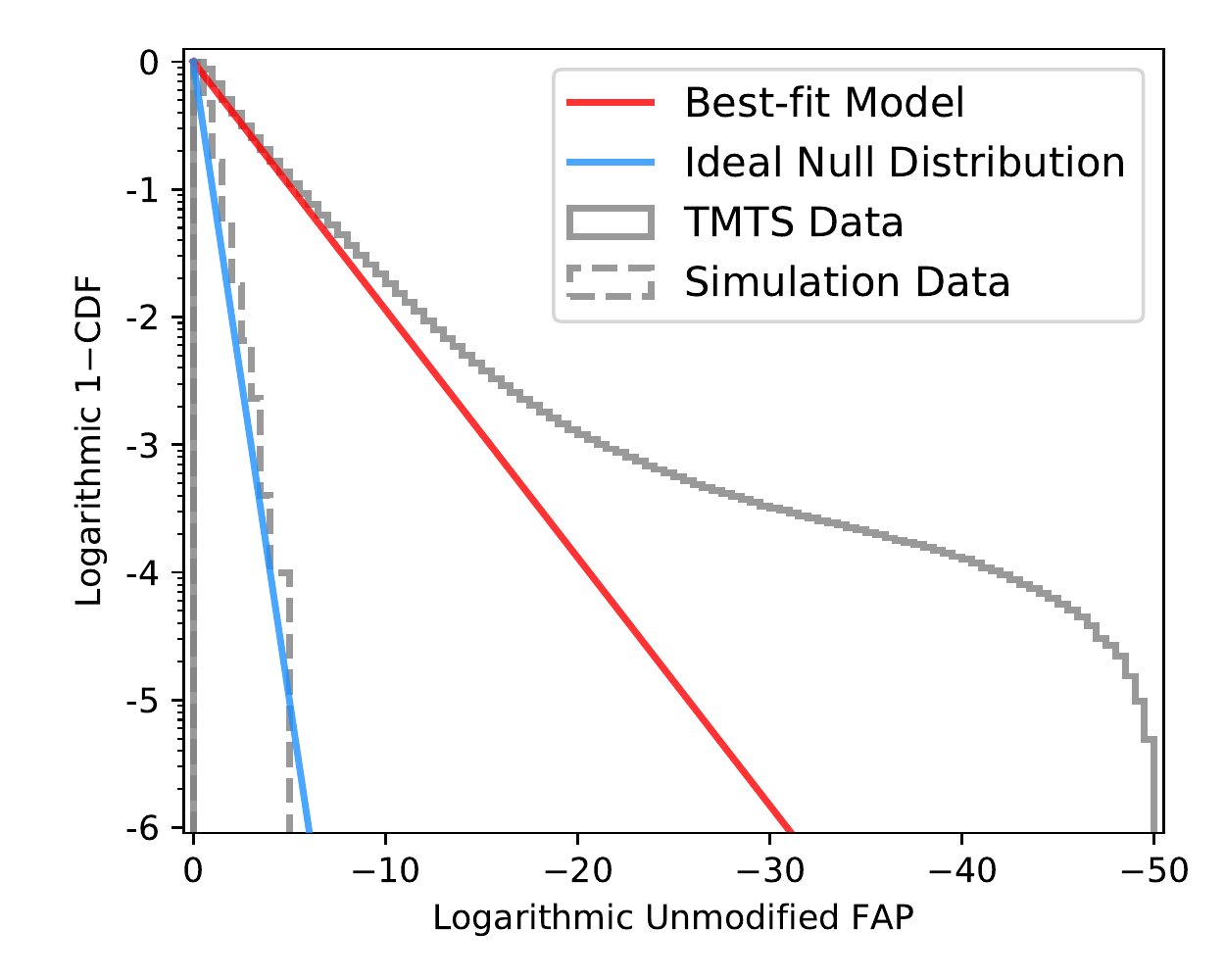}
    \caption{
    \textbf{
    Cumulative distribution function (CDF) of unmodified false alarm probability (FAP) for periodicity detection in the \tmtsdata dataset.
     } The bin size is 0.5.
The grey solid line represents the FAPs obtained from the \tmtsdata light curves, and the red solid line is the model fitted to the grey bins above 1$-$CDF$=0.1$.
The grey dashed line represents the FAP distribution derived from 10,000 simulated time series obeying a normal distribution, and the blue solid line is the ideal null distribution with 1$-$CDF$=$FAP.
     }
    \label{fig:periodic_prob}
\end{figure}

Due to potential nonuniform sampling caused by some ``bad'' measurements, we test the periodicity of TMTS light curves using the Lomb–Scargle periodogram (LSP hereafter, \citealt{Lomb+1976,Scargle+1982,VanderPlas+2018+LSP_understanding} ). The LSP here is defined as
\begin{equation}
\begin{aligned}
P(f)=\frac{1}{2 \sigma^2}\times \left \{ \frac{ \left [\sum\limits_{i=1}^{N} F_{i} \times \cos\, 2\pi f(t_i-\tau) \right ]^2}{\sum\limits_{i=1}^{N} \cos^2\, 2\pi f(t_i-\tau)}  \right . \\
  +\left . \frac{ \left [\sum\limits_{i=1}^{N} F_{i} \times \sin\, 2\pi f(t_i-\tau) \right ]^2}{\sum\limits_{i=1}^{N} \sin^2\, 2\pi f(t_i-\tau)}   \right \}, \\
 {\rm and}\,\, \tau= \frac{1}{4 \pi f} \times  \left ( \arctan \, \frac{\sum\limits_{i=1}^{N} \sin\, 4\pi f t_i  }{\sum\limits_{i=1}^{N} \cos\, 4\pi f t_i } \right )  \\
\end{aligned}
\end{equation}
where $F_{i}$ is the flux at epoch $t_i$ after the calibration (see Section~\ref{sec:reduction}), $f$ is the test frequency and $\sigma^2$ is the variance of the fluxes. 
The LSP here is normalized by the variance and thus the white noise in LSP follows the exponential distribution as $\exp{(-z)}$ (see also \citealt{Coughlin+etal+2020+ZTFprojectionII}). We determine the most likely photometric period by searching the highest LSP peak $P_{\rm max}$ in the frequency range of $ 3/2T \leq f \leq { f_{\rm nyq}} $,  where $T$ is the time span of the observations and $f_{\rm nyq}$ is the (pseudo-)Nyquist frequency \citep{VanderPlas+2018+LSP_understanding}, which can be estimated as a half of average sampling rate ($\approx 1/75$~Hz). Notice that, the observations must cover one and half cycles before we can determine its periodicity.

\begin{figure*}
    \includegraphics[width=0.94\textwidth]{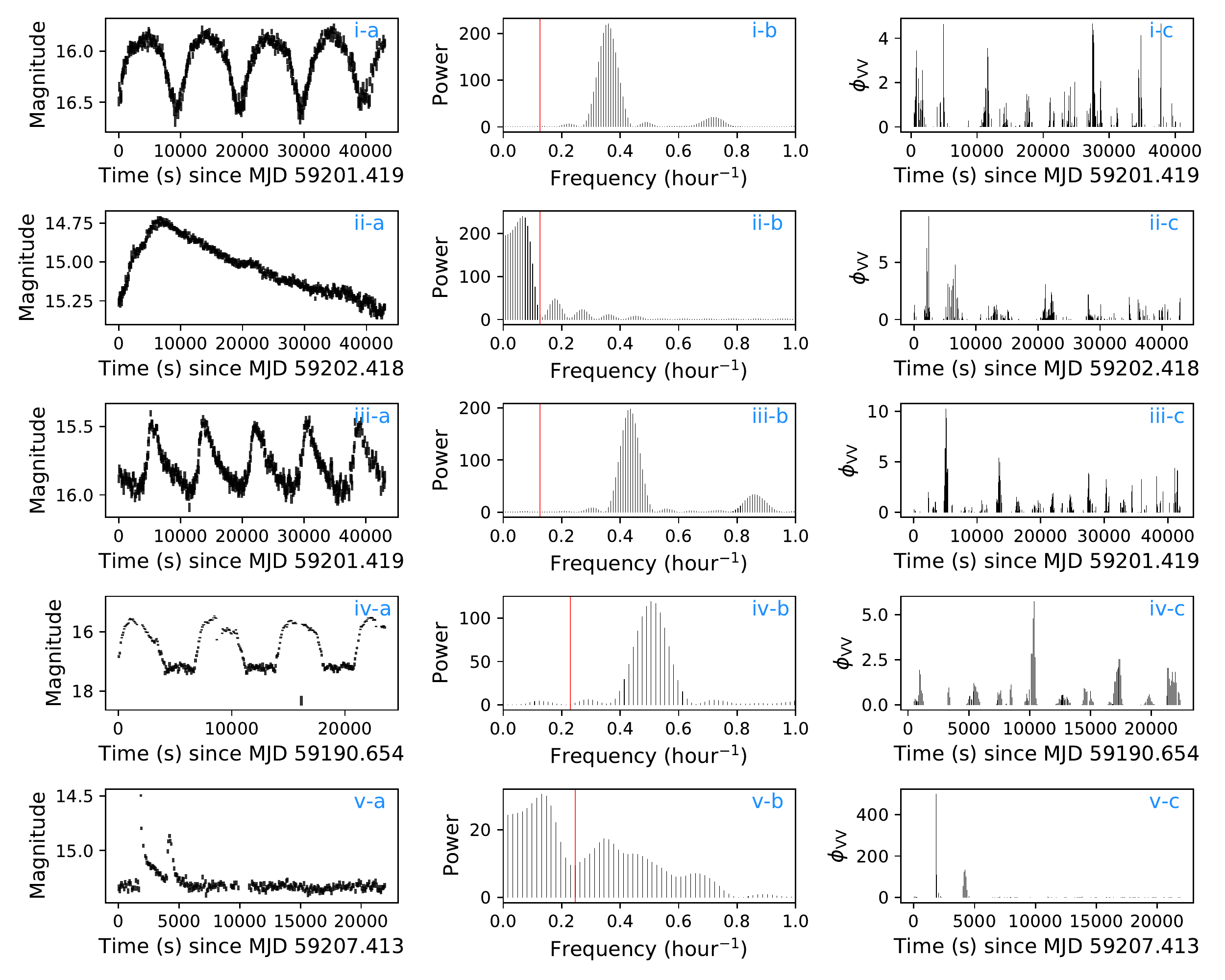}
    \caption{
    \textbf{
    Examples for detections of periodicity variables and flare stars.}
    The row \emph{i} to \emph{v} correspond to a W UMa-type eclipsing binary, a RR Lyr-type variable, a $\delta$ Scuti star, an AM Her-type cataclysmic variable and a multi-peak flare, respectively.
    The column \emph{a}, \emph{b} and \emph{c} represent the TMTS light curves, Lomb-Scargle periodograms, and time series of $\phi_{\rm VV}$, respectively. 
    The red lines in column \emph{b} represent the frequency threshold (i.e., $3/2T$) used to investigate the red noise in Lomb-Scargle periodograms. { Notice that, the Lomb-Scargle periodograms here are a zoom-in version, the complete Lomb-Scargle periodograms cover the frequency even higher than 20~hour$^{-1}$. }
     }
    \label{fig:lsp_and_phivv}
\end{figure*}

Based on the cumulative distribution function (CDF) of $\exp{(-z)}$ and the multiplicative property of the independent probabilities, the false alarm probability (FAP, see \citealt{Lomb+1976}) of periodicity can be estimated as 
\begin{equation}
{\rm FAP}=1-[1-\exp{({ -P_{\rm max} })}]^{N_{\rm eff}}
\label{Eq+FAP}
\end{equation}
where $N_{\rm eff}$ is the number of independent frequencies, which can be calculated by $N_{\rm eff}=f_{\rm nyq}T$ in approximation \citep{VanderPlas+2018+LSP_understanding}, namely a half of total number of epochs.
Notice that the estimate of FAP is completely dependent on the assumption of white noise. However, LSP power tends to be higher at lower frequency because of the effect of red noise, thus the resultant FAP could be seriously underestimated, especially at the low-frequency end.
Due to relatively short duration of our continuous photometry, the LSP is more likely to be polluted by the red noise generated from non-periodic or long-period variations. 
Therefore, we also search for the high LSP powers at lower frequency range (i.e. $f < 3/2T $), since strong powers in the low-frequency range (as $P^{\rm red}$) are very likely caused by red noise rather than real periodic behavior. All light curves with $P^{\rm red} > P_{\rm max} $ were marked to indicate the possible red noises in the LSPs.
Several samples of period search are shown in Figure~\ref{fig:lsp_and_phivv}. Noted that both panel \emph{ii-b} and panel \emph{v-b} have { higher powers $P^{\rm red}$} below the frequency threshold, implying that they suffered non-periodic variations during the observations.

For the purpose of checking the periodicity FAPs obtained from the light curves in the \tmtsdata dataset, we plotted their cumulative distribution function (CDF) in Figure~\ref{fig:periodic_prob}. For comparison, we also generated 10,000 simulated time series, and each is composed of random 100--1000 points that obey a normal distribution. 
As shown in Figure~\ref{fig:periodic_prob}, the periodicity FAPs, calculated from these simulated time series, follows exactly the ideal null distribution 1$-$CDF$=$FAP. This implies that the method of obtaining periodicity FAPs is feasible if the TMTS measurements are independent and follow a Gaussian distribution.
However, our FAP, estimated from real dataset, deviates significantly from the ideal null distribution. Such a deviation was also found to exist in the dataset of other survey mission \citep{Drake+etal+2013,Drake+etal+2014+catalina_period}.
It is known that the detectable periodic variable stars take only a small percentage of all observed sources \citep{Drake+etal+2014+catalina_period,Drake+etal+2017+catalina_period,Chen+etal+2020+ZTF_periodic,ZTF+DR1+2020},
the periodicity FAP discussed here can be seriously underestimated in practise. 

\begin{figure}
    \includegraphics[width=0.47\textwidth]{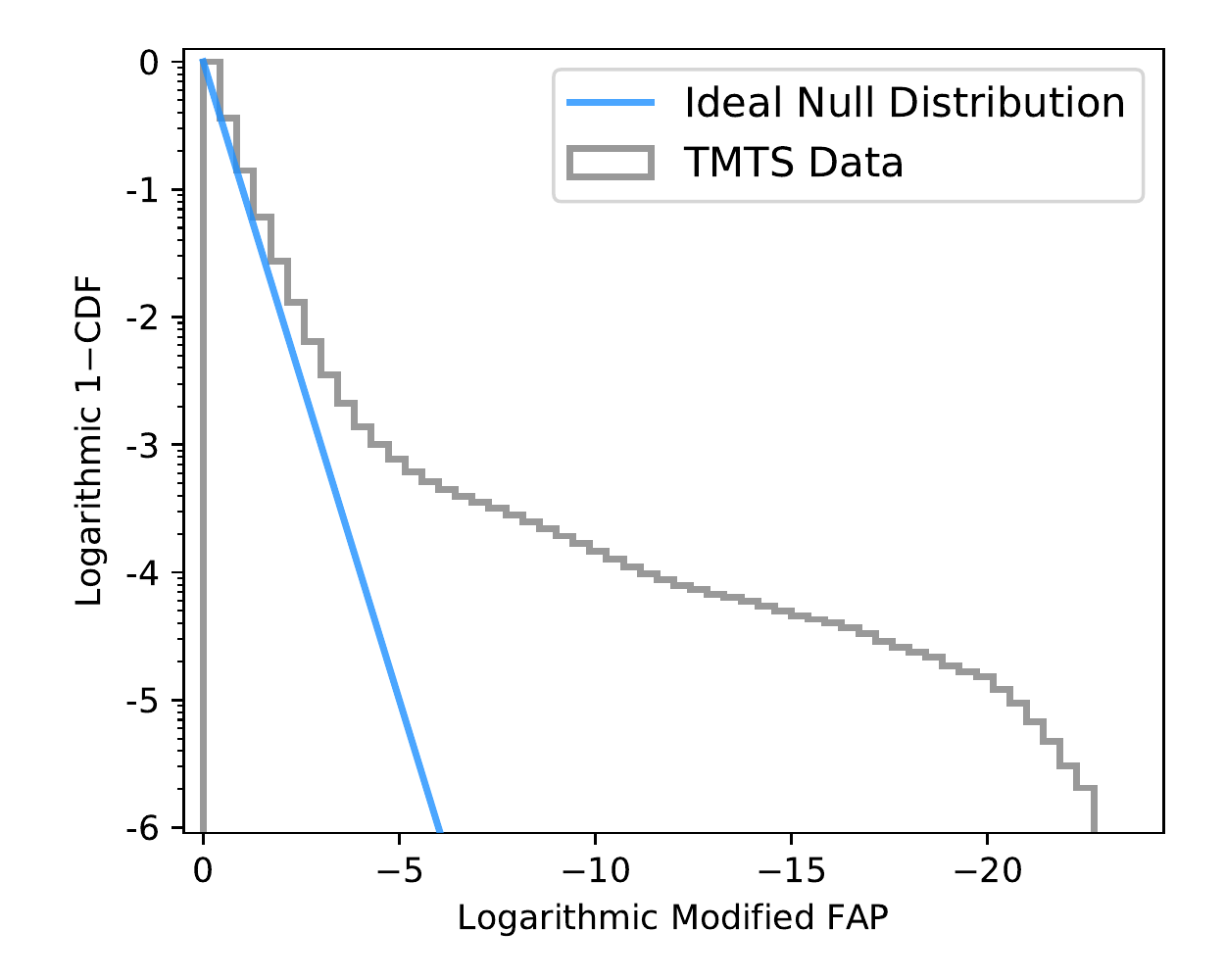}
    \caption{
    \textbf{
    Cumulative distribution function (CDF) of modified false alarm probability (FAP) for periodicity detection in the \tmtsdata dataset.
     } The bin size is 0.5.
The grey solid line represents the FAP modified by $k=0.225$,
and the blue solid line is the ideal null distribution with 1$-$CDF$=$FAP.
     }
    \label{fig:modified_periodic_prob}
\end{figure}

In order to derive more reliable estimate of periodicity FAPs, some methods have been developed, such as the \emph{Baluev}'s method \citep{Baluev+2008+FAP_corrected} and bootstrap method \citep{Ivezi+2014+book}. As an alternative, we construct a true null distribution by real dataset since we have already calculated the FAPs for all available light curves of the \tmtsdata { dataset}. By assuming that most observed sources are non-periodic sources (typically $\gtrsim 90\%$ ) and periodic sources tend to have higher LSP peaks and thus lower FAPs, we can take { the highest 90\% FAPs as approximated} null samples. We found these null samples (corresponding to the 1$-$CDF$> 0.1$ part of grey solid line in Figure~\ref{fig:periodic_prob}) follow a straight line in the logarithmic 1$-$CDF versus logarithmic FAP diagram, implying that the null distribution of real data may differ from the ideal null distribution by only a constant $k$, namely $\log_{10}(1-{\rm CDF})=k\times\log_{10} {\rm FAP}$. By fitting the distribution of these null samples, we obtained $k=0.225$ for all light curves of the \tmtsdata { dataset} and then the modified false alarm probabilities can be expressed as $ {\rm FAP_{mod} }= {\rm FAP}^k$. 
Notice that the $k$ value actually varies for different datasets as the performance of flux measurements is dependent on the observation conditions. Thus, a \tmtslocal $k$ value is determined from daily observation dataset in real time and  \tmtslocal modified FAPs are generated for the light curves.

The distribution of modified FAP is shown in Figure~\ref{fig:modified_periodic_prob}, where one can see that the area between the grey line and blue line corresponds to the candidates of periodic variable.
 About $99\%$ TMTS light curves [i.e. $\log_{10}(1-{\rm CDF}) >-2$] match the ideal null distribution, implying that the (detectable) periodic light curves account for only a few thousandths of \tmtsdata dataset. 
 { Notice that, due to the limitation of current observation duration (typically within a night), it is difficult to reveal long-period variables (e.g. $P>0.5$~day, see details in Section~\ref{sec:periodic_variable_sources} ) using the \tmtsdata dataset. But the number of periodic variables will be greatly improved with the ongoing of supernova survey of TMTS.
 }

\subsection{Flare search}

A common method of flare detection is to search for outliers in the light curves for which non-flare variations (e.g. large-amplitude and long-duration variations ) have been removed. For the purpose of avoiding false flares caused by instrumental errors or cosmic rays, the flare search often requires at least two consecutive outliers rather than one-point outliers \citep{Walkowicz+etal+2011+flares,Osten+etal+2012+flare_archive,Yang+etal+2018+SC_LC_flares}.
In order to remove the non-flare variations, \cite{Osten+etal+2012+flare_archive} use two different models to fit periodic and non-periodic light curves, respectively. To avoid the comparison of goodness for two models and speed up the data analysis, we fit all light curves by a unified compound model of 4th-order Fourier series \citep{Pojmanski+2002+06hvariable, Kim+etal+2016+UPSILON,Drake+etal+2014+catalina_period,Drake+etal+2017+catalina_period} and 2nd-order polynomial.
The polynomial terms here are used to offset the potential long-scale variations.
Notice that the purpose of fitting here is to remove the non-flare variations, rather than modeling the true variations of light curves. The compound model is expressed as
\begin{equation}
F_{i}^{\rm model}=\sum\limits_{j=0}^2 c_i\times t_i^{j} + \sum\limits_{k=1}^4 a_k\times \cos (2\pi \,k\,f_{\rm max}\,t_i ) + b_k\times \sin (2\pi  \,k\,f_{\rm max}\, t_i )
\end{equation}
where $f_{\rm max}$ is the frequency corresponding to the highest power $P_{\rm max}$ in the LSP (see Section~\ref{sec:periodicity_detection}). Notice that even if a light curve is non-periodic, we can still find a $f_{\rm max}$ in its LSP. For such a light curve, our model is still applicable while the values of best-fit $a_k$ and $b_k$ are very small.

The residual flux $F_{i}^{\rm res}= F_{i} - F_{i}^{\rm model}$ can be easily obtained and the normalized residual could be calculated as
\begin{equation}
r_i= \frac{F_{i}^{\rm res}- \overline{F^{\rm res}}}{\sigma^{\rm res}}
\end{equation}
where $\overline{F^{\rm res}}$ and $\sigma^{\rm res}$ represent the median flux and the robust standard deviation  of residual fluxes, respectively. 
It is worth noting that, we adopted the robust StD instead of the uncertainty in flux measurements, since the former can reveal true scatter in the residual fluxes. 
In this way, the normalised residuals (except the points corresponding to flares) will obey a normal distribution, which is the prerequisite to estimate the significance of selected flare candidates.

The flare candidates are selected by locating the maximum $\phi_{\rm VV}$ in a time series  \citep{Osten+etal+2012+flare_archive}, where the $\phi_{\rm VV}$ is defined as the product of continuous two normalized residuals, 
\begin{equation}
 \phi_{\rm VV, i}=r_{i} \times r_{i+1}
\end{equation}
The examples of flare search using time series of $\phi_{\rm VV}$ are also shown in Figure~\ref{fig:lsp_and_phivv}, where one can see that only the row \emph{v} (corresponding to a flare ) has very strong $\phi_{\rm VV}$ values in its time series.

To estimate the false discovery rate (FDR), some studies \citep{Kowalski+etal+2009+FDR_flare,Osten+etal+2012+flare_archive,Paudel+etal+2018+K2_flare,Paudel+etal+2020_LDwarf_flares} have applied the FDR analysis following \cite{Miller+etal+2001+FDR}. However, the \emph{Miller}'s method is not applied to our project,  because it requires a large number of null samples manually selected from the dataset of light curves, but this work cannot be finished automatically in real time by our pipeline. Therefore, we explored the mathematical formula for the null distribution of $\phi_{\rm VV}$.

\begin{figure}
    \includegraphics[width=0.47\textwidth]{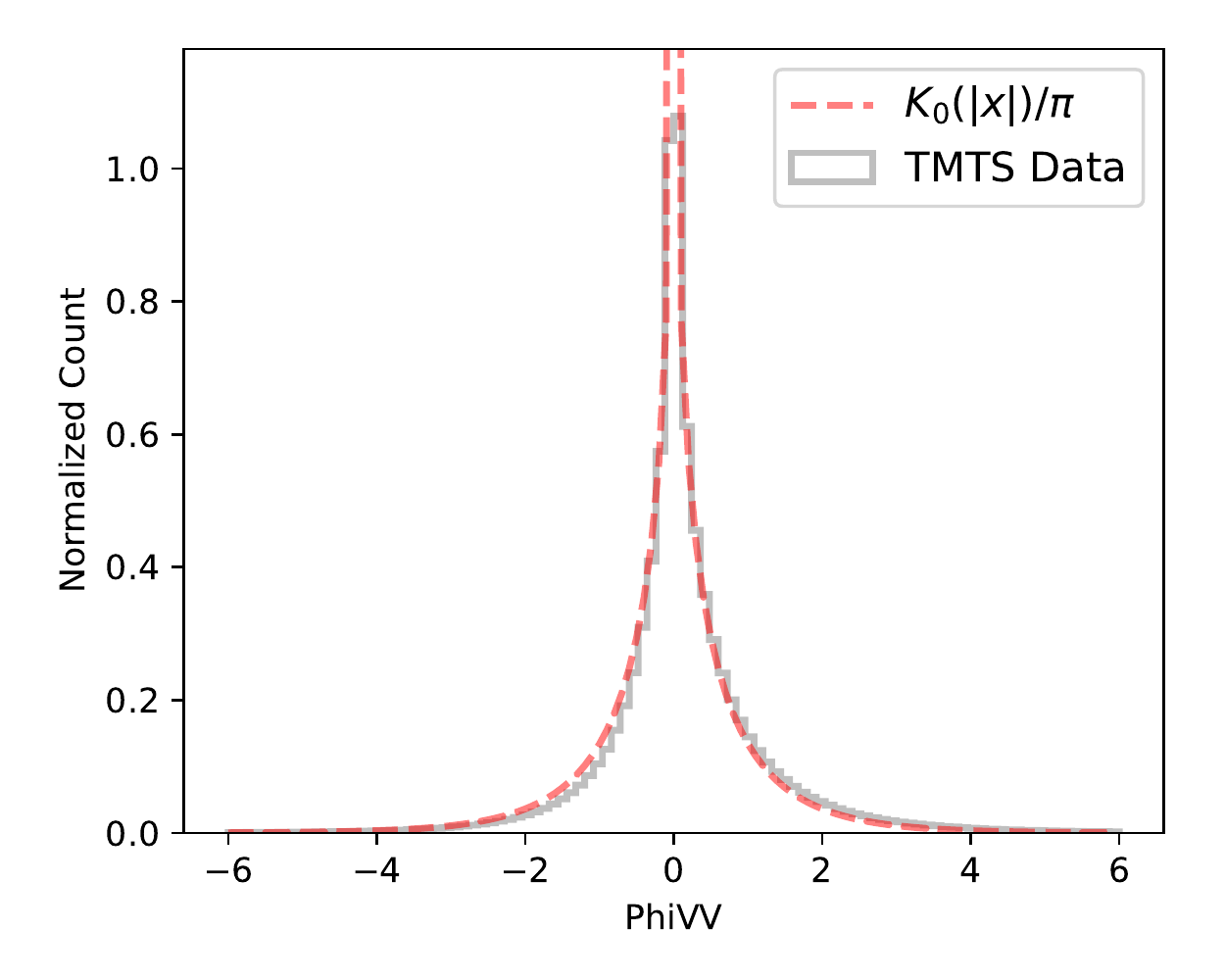}
    \caption{
    \textbf{
    The density distribution of $\phi_{\rm VV}$ calculated from the TMTS observation on December 19, 2020.
     }
     The observation data include about 36~million $\phi_{\rm VV}$ measurements from light curves of about 82 thousand sources.
     The bin size is 0.06.
     }
    \label{fig:phivv_distribution}
\end{figure}

As introduced above, the (non-flare) normalized residual fluxes obey a normal distribution. By assuming successive residuals are independent of each other, the product of two normalized residuals (i.e. $\phi_{\rm VV}$) should follow the probability density function (PDF) as indicated below,
\begin{equation}
{\rm PDF}= \frac{K_0(\vert \phi_{\rm VV}  \vert)}{\pi}= \frac{1}{\pi} \int^{\infty}_{0} \frac{\cos (\phi_{\rm VV}\,t)}{\sqrt{t^2+1}}\,{\rm d}\,t
\end{equation}
where $K_0$ is the special ($n=0$) case of \emph{modified Bessel function of the second kind} \citep{Abramowitz+Stegun+1972+mathematical+function}.
Figure~\ref{fig:phivv_distribution} shows the density distribution of for about 36~million $\phi_{\rm VV}$ measurements from the TMTS observation conducted on December 19, 2020.
The Bessel function is characterized by the density distribution of $\phi_{\rm VV}$, implying that the CDF of the Bessel function can be applied to estimate the FDR for flare candidates.

Similar to Eq~\ref{Eq+FAP}, the FDR of flares can be estimated as 
\begin{equation}
{\rm FDR}=1-[1 - \frac{1}{2\pi} \int^{\infty}_{ \phi_{\rm VV, max}  } \,K_0(\vert x \vert) \,{\rm d}\,x    ]^{N-1}
\label{Eq+FDR}
\end{equation}
where $\phi_{\rm VV, max}$ is the maximum value of $\phi_{\rm VV}$ inferred in a time series and
 $N-1$ is the number of $\phi_{\rm VV}$. Notice that, for the purpose of selecting flares,
we must exclude the $\phi_{\rm VV}$ derived from the product of a pair of negative $r_i$ values (thus the integral probability term is multiplied by a factor of $\frac{1}{2}$ in Eq~\ref{Eq+FDR}).

\begin{figure}
    \includegraphics[width=0.46\textwidth]{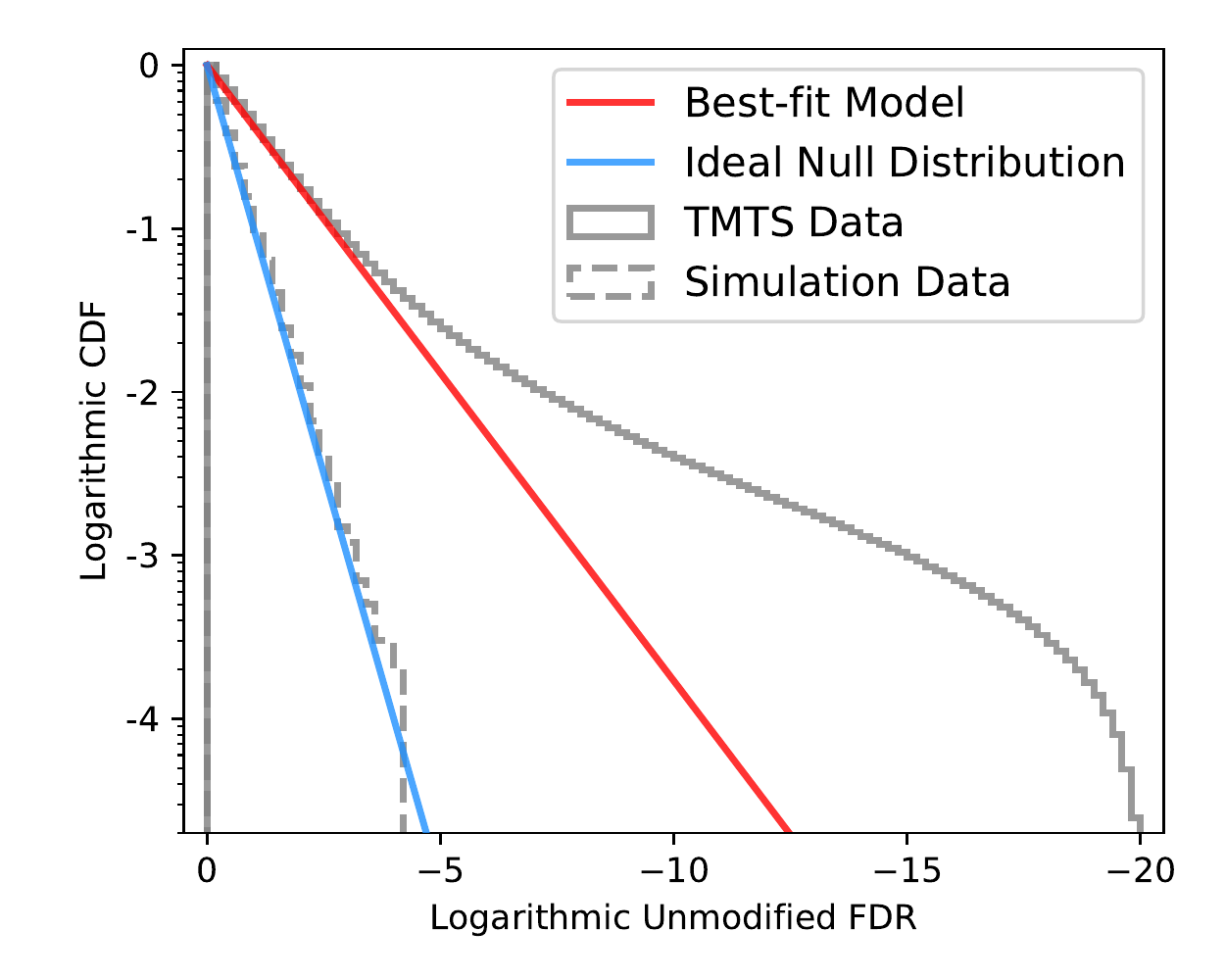}
    \caption{
    \textbf{
    Cumulative distribution function (CDF) of unmodified false discovery rate (FDR) for flare detection in the \tmtsdata light curves.
     }
     The bin size is 0.2.
     }
    \label{fig:phivv_prob}
\end{figure}

The cumulative distribution function of flare FDRs for about 4.9~million { light curves of \tmtsdata dataset} is shown in Figure~\ref{fig:phivv_prob}. The FDR distribution generated from 10,000 simulated random time series (see more in Section~\ref{sec:periodicity_detection}) matches well with the ideal null distribution, suggesting that our mathematical formula of calculating the FDR is applicable for normally distributed and independent residuals. However, as the successive normalized residuals are not completely independent of each other, the FDRs generated from the \tmtsdata dataset deviate significantly from that of the ideal null distribution. Here we applied the same methods and assumptions described in Section~\ref{sec:periodicity_detection} to modify the FDR. By fitting the cumulative distribution above 0.1, we obtained $k=0.376$ for the overall \tmtsdata data and modified the false discovery rate by ${\rm FDR}_{\rm mod}={\rm FDR}^k$.
On the other hand, for daily observation dataset, our pipeline of data analysis will also automatically determine a \tmtslocal $k$ value, which helps modify the FDR derived from daily observation data in a more accurate way.

\subsection{Cross match with other catalogs}
In order to obtain additional information (e.g. distance and radial velocity) for source selections, 
we have cross-matched all of the TMTS sources with the Gaia DR2 database and the LAMOST DR7 (including both low- and medium-resolution spectra) catalogs.

\subsubsection{Gaia}

The Gaia DR2 covers 1.69 billion sources brighter than 21 mag, among which astrometric positions, parallax and proper motion parameters are available for more than 1.33 billion sources \citep{Gaia_collaboration+2018+data}. Most of these sources also have photometric data in G (330--1050~nm), B (330--630~nm) and R bands (630--1050~nm).

Out of the 4.87 million light curves from the \tmtsdata dataset,  
about 4.83 million light curves (99\% ) are found to have the Gaia-DR2 counterparts.
The remained sources without Gaia counterparts are either transients or bad detections.
After removing a part of sources that are located in multiple LAMOST plates or repeatedly observed by different telescopes of TMTS, the total number of the Gaia-DR2 counterparts is 4.26 million.
Among them, about  4.18 million (98\%) Gaia sources have parallax measurements,
but only 2.63 million (62\%) have reliable parallax measurements (i.e. $\sigma_\varpi/\varpi \leq 0.2$ here).

\begin{figure}
    \includegraphics[width=0.47\textwidth]{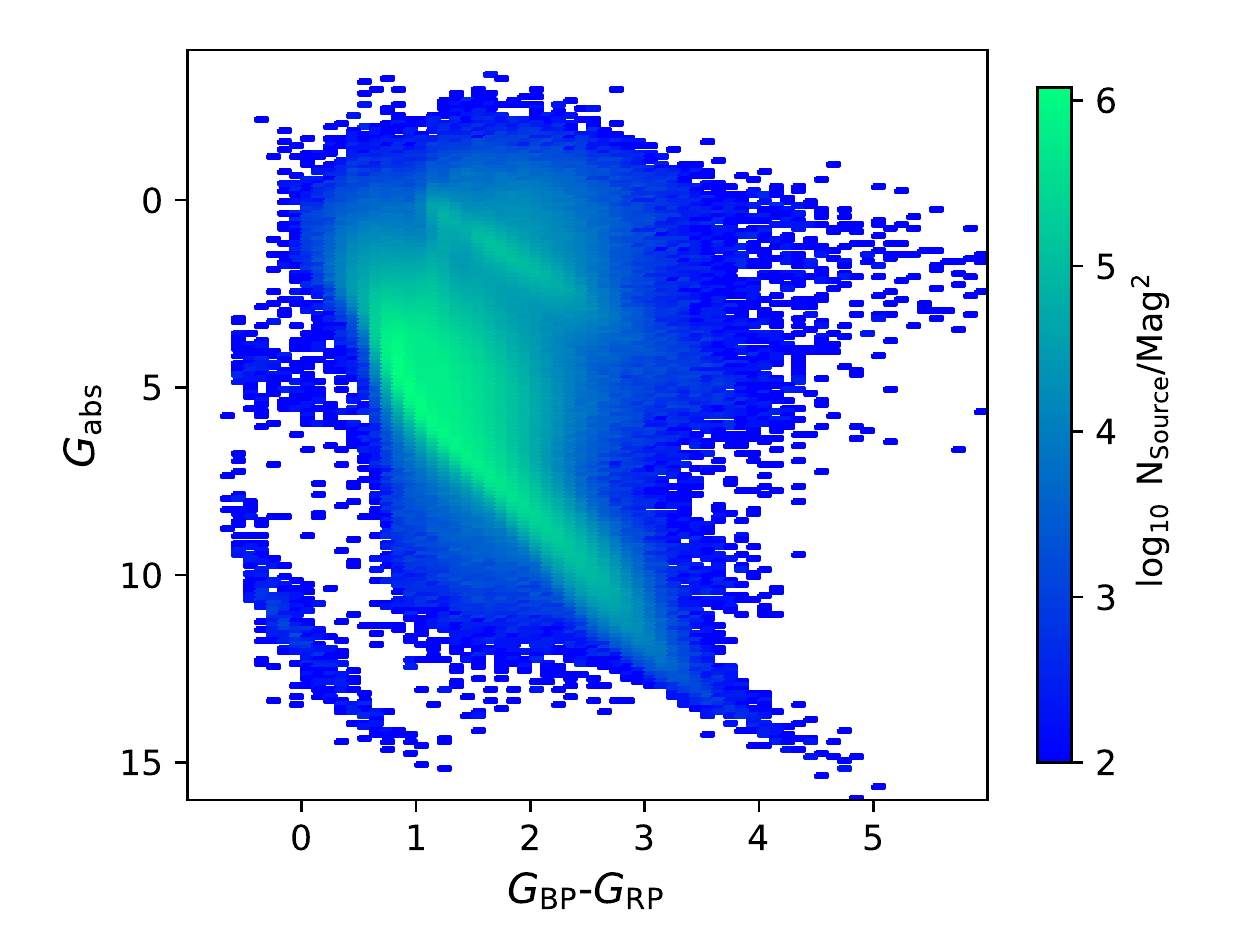}
    \caption{
    \textbf{
Density distribution of the Gaia DR2-TMTS sources across the Hertzsprung–Russell (HR) diagram.} The bins size is 0.1$\times$0.1 mag$^2$, and the total number of sources is 2.63 million.
     }
    \label{fig:gaia_hr}
\end{figure}

Based on parallax, G magnitude and color ($G_{\rm BP}-G_{\rm RP}$), we can plot the Hertzsprung-Russell (HR) diagram for these Gaia DR2-TMTS sources, which is a very valid tool to select the white dwarfs \citep{Esteban+etal+2018+gaia_wd,Fusillo+etal+2019+gaia_wd,Pelisoli+Joris+2019+ELM,Kim+etal+2020+wd} and hot subdwarf stars \citep{Geier+etal+2019+hot_subdwarf,Geier+2020+subdwarf}. 
As shown in Figure~\ref{fig:gaia_hr} , the Gaia DR2-TMTS sources cover a wide area across the HR diagram.
The reddening corrections were not applied to these sources.
The high-density areas in HR diagram correspond to giant stars and main-sequence stars, which include some classes of variables, such as pulsating stars and eclipsing binaries.
The TMTS has also captured a small number of white dwarfs in its first-year observations. By applying a simple set of cuts for $G_{\rm abs}$ and $G_{\rm BP}-G_{\rm RP}$ (the Eq~2--5 in \citealt{Fusillo+etal+2019+gaia_wd} ), we identified 565 ($\approx$0.02\%) white dwarf candidates out of the 2.63 million TMTS sources.
Furthermore, we can identify cataclysmic variables (CVs, \citealt{Pala+etal+2020+gaia_CV}) and  $\delta$ Scuti stars in periodic variables (see details in Section~\ref{sec:periodic_variable_sources}).

\subsubsection{Lamost}
From 2011 to 2019, the LAMOST spectroscopic survey has obtained 10.6 million low-resolution ($R\sim 1800$) spectra and 11.4 million single-exposure medium-resolution ($R\sim 7500$) spectra for about 8.9 million LAMOST targets brighter than 17.5 mag (see more in \citealt{Cui+etal+2012+LAMOST,Zhao+etal+2012+LAMOST} and \url{http://dr7.lamost.org}).
In the LAMOST DR7 catalog \emph{v1.2}, all of the LAMOST targets have already been cross-matched with that of the Gaia DR2 catalog.

Plenty of atmospheric parameters (including effective temperature, surface gravity and metallicity) can be inferred from the observed spectra  by applying LAMOST Stellar Parameter pipeline (LASP).
The abundant spectral parameters from LAMOST can help identify the variable stars from the TMTS.
It is worth noting that, LAMOST can provide radial velocity (RV) measurements for millions of stars. 
These RV measurements were determined by LASP or cross correlation with KURUCZ synthetic templates \citep{Wang+etal+2019+RV_measurements}.
As multi-epoch spectra are available for millions of objects, the LAMOST data can reveal radial velocity variations for a large number of stars.  Hence, the LAMOST data can be used to select those with significant RV variations
(e.g. eclipsing binaries, \citealt{Yang+etal+2020+EB_lamost} ) and even the binaries  harboring an invisible high-mass companion (i.e. black hole binary candidates, \citealt{Thompson+etal+2019+Science,Liu+etal+2019+Nature,Yi+etal+2019+theory_bh,Zheng+etal+2019+lamost_bh}  ).

\begin{figure}
    \includegraphics[width=0.47\textwidth]{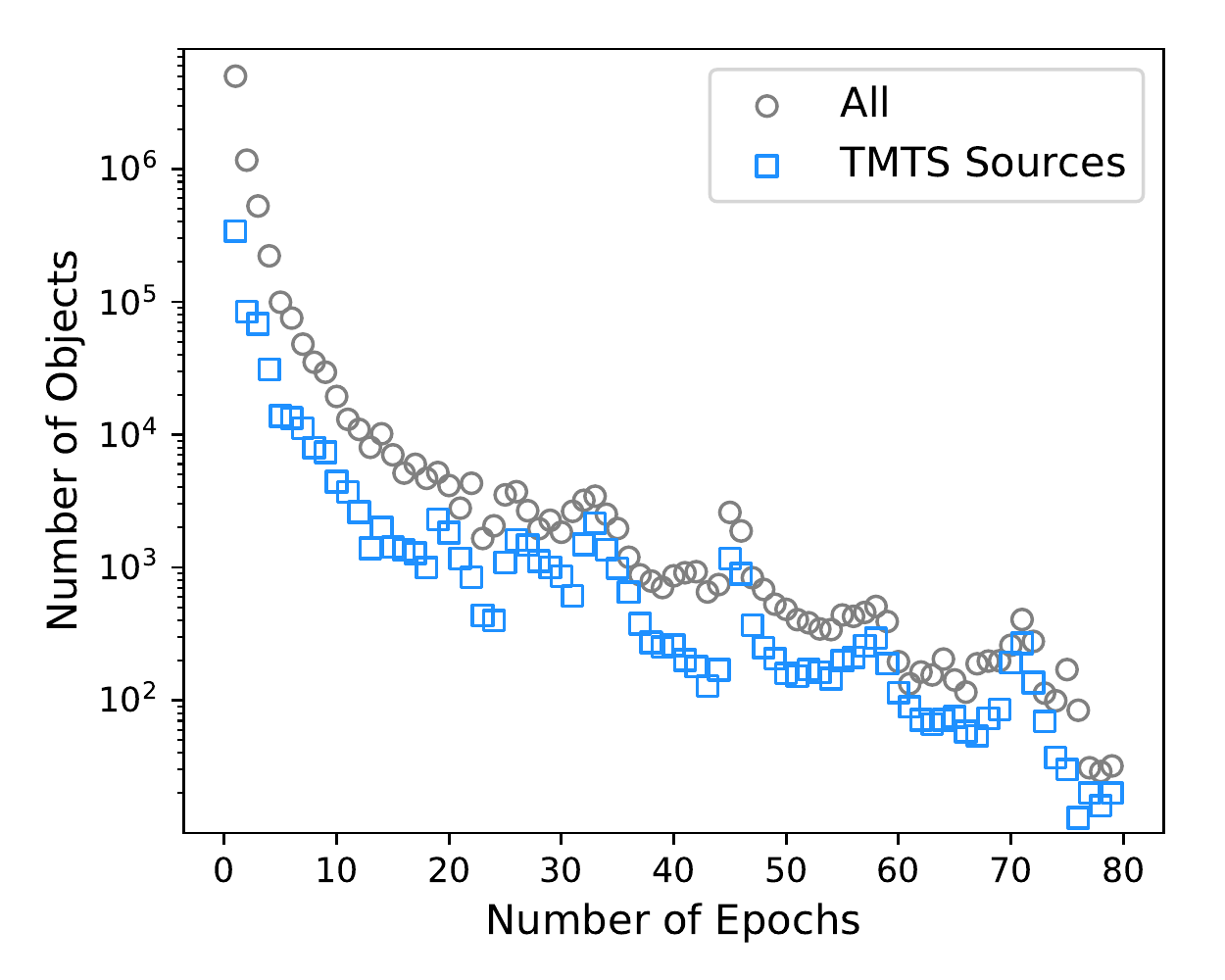}
    \caption{
    \textbf{Number distribution of LAMOST DR7 epochs (including both low-resolution and single-exposure medium-resolution spectra) .}
    The blue squares represents the distribution for the LAMOST-TMTS sources.
     } 
    \label{fig:n_exposure_dist}
\end{figure}

Figure~\ref{fig:n_exposure_dist} shows the number of objects corresponding to the LAMOST DR7 spectra versus the number of the corresponding spectra which contain both low-resolution and single-exposure medium-resolution spectra. 
About 2.35 million sources ($32\%$ of LAMOST sources) have 2 to over 70 epochs, and the number of sources decrease with the number of epochs.
After cross-matching the TMTS sources in \tmtsdata dataset with the LAMOST DR7 catalog, we find that there are {626 thousand sources (8.5 \% of the LAMOST sources)} in common. Among them, 285 thousand TMTS sources have multi-epoch spectra. For a five-year survey, the TMTS is expected to provide high-cadence photometric data for more than half of the LAMOST sources.

\section{Preliminary Results}

\label{results}
In this section, we present some preliminary results from the first-year high-cadence photometric observations of the LAMOST fields with the TMTS.
The result of searching for periodic variable sources is shown in Section~\ref{sec:periodic_variable_sources}, and the result for flares is shown in Section~\ref{sec:flare_stars}.

\subsection{Periodic Variable Sources}

\label{sec:periodic_variable_sources}

\begin{figure}
    \includegraphics[width=0.47\textwidth]{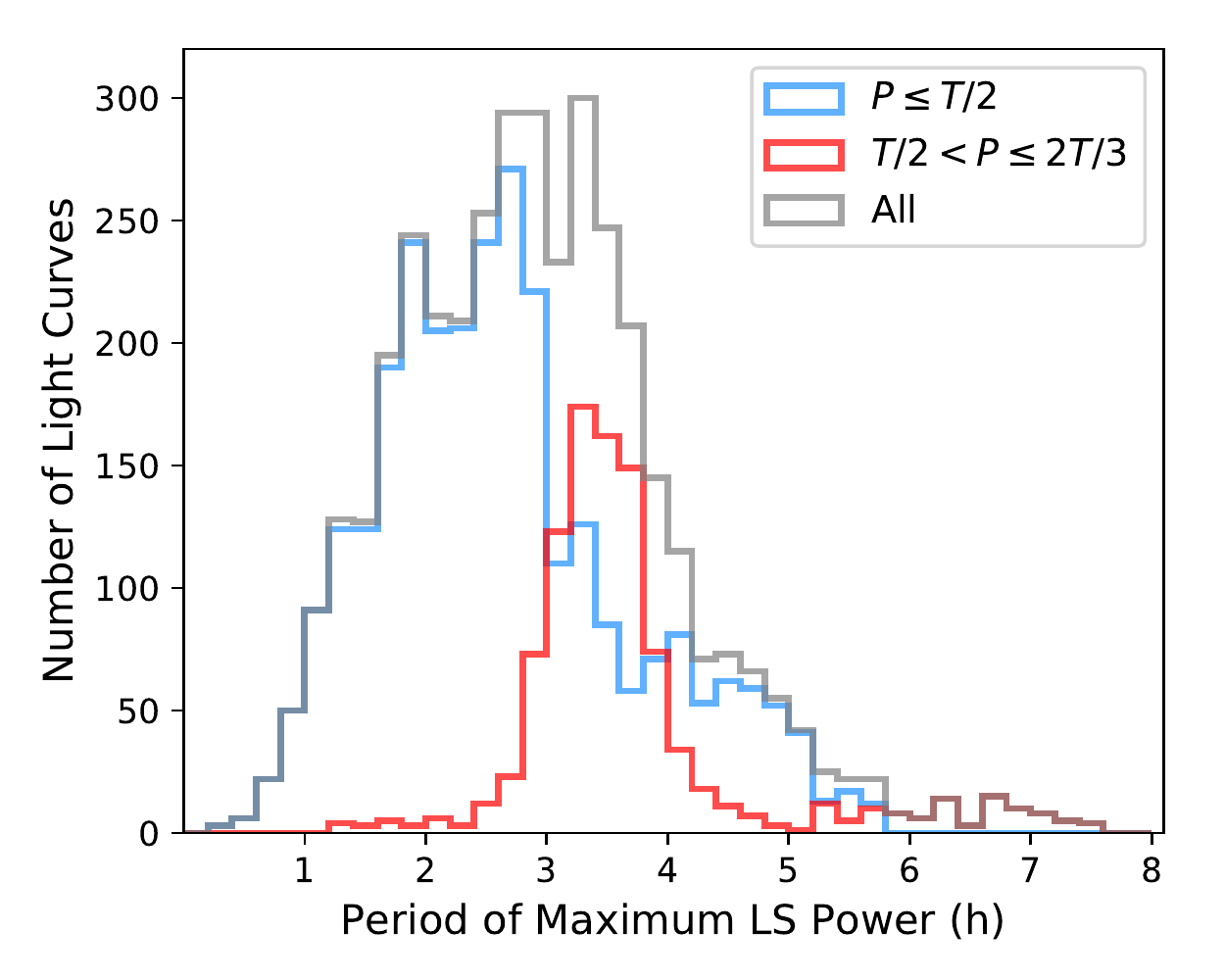}
    \caption{
    \textbf{  Period distribution of selected light curves from the \tmtsdata dataset.}
    The red and blue lines represent the sets of light curves selected with different ratios of period of light variation to observation duration.
     } 
    \label{fig:periodic_variables_dist}
\end{figure}

\begin{figure}
    \includegraphics[width=0.47\textwidth]{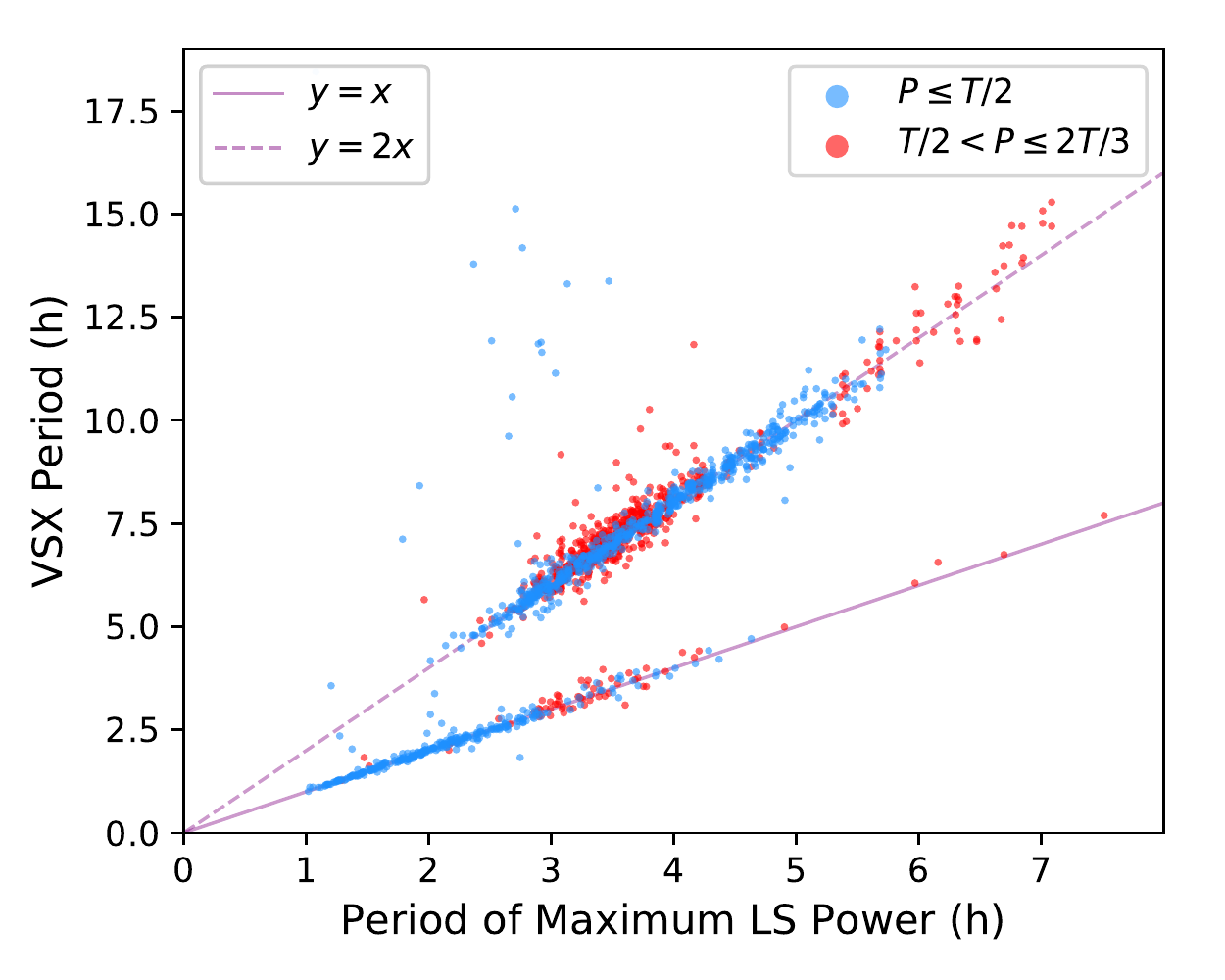}
    \caption{
    \textbf{Period of the highest peak in the periodogram from the \tmtsdata dataset versus the period given by the International Variable Star Index (VSX).}
    The solid and dashed lines represent the relation $y=x$ and $y=2x$, respectively.
     } 
    \label{fig:period_vsx}
\end{figure}

\begin{figure}
    \includegraphics[width=0.47\textwidth]{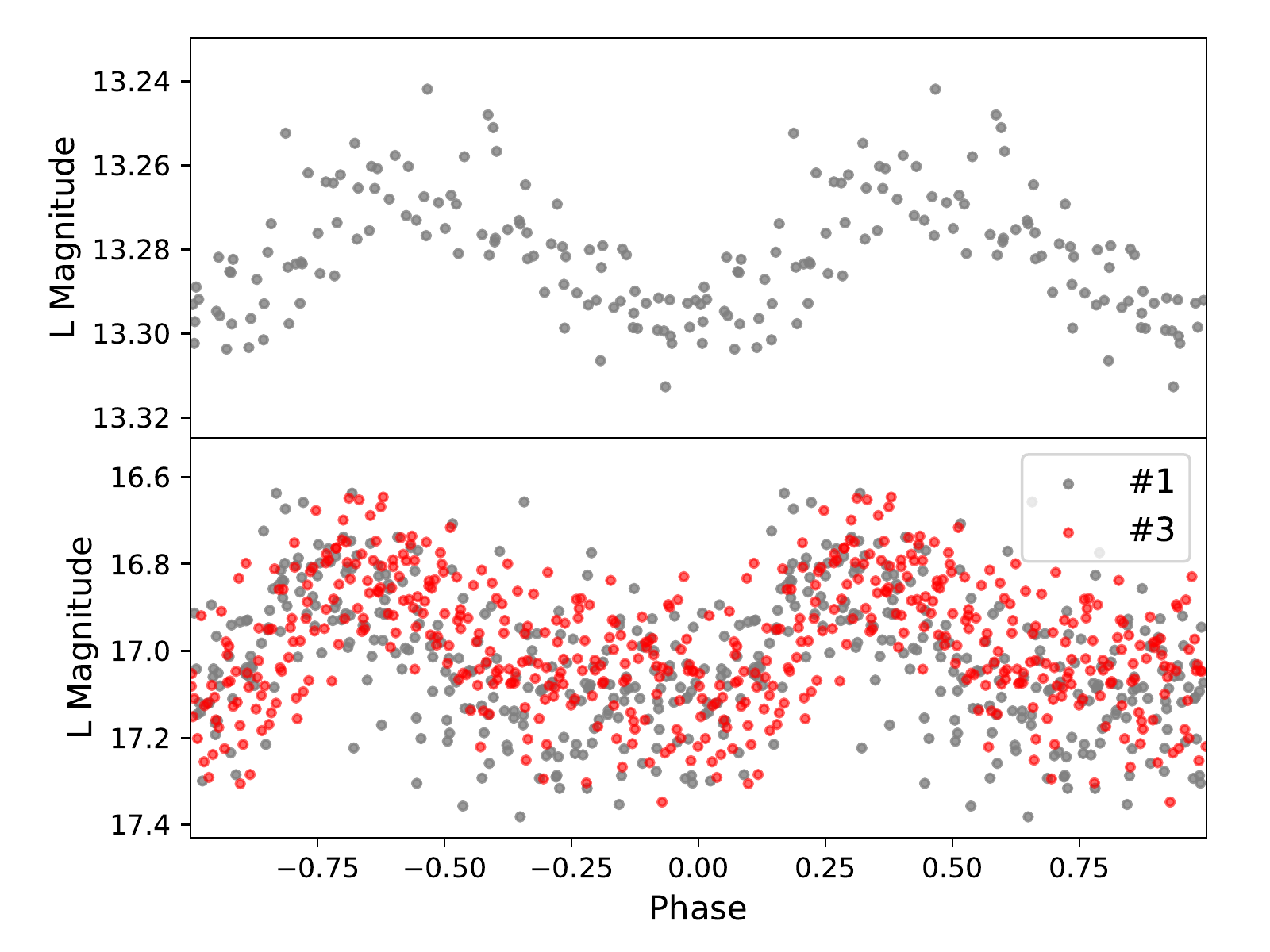}
    \caption{ 
    \textbf{The phase-folded light curves for two shortest-period stars from \tmtsdata.} 
    The upper panel shows a $\delta$ Scuti star candidate with a period of 18.4~minutes and an amplitude of 0.03~mag; the lower panel shows a blue large-amplitude pulsator (BLAP) candidate with a period of 18.9 minutes and an amplitude of 0.26~mag. Notice that, the BLAP candidate has been observed by telescope \#1 (grey) and \#3 (red) within different nights, respectively.
     }
    \label{fig:phased_lc_examples}
\end{figure}

\begin{figure*}
    \includegraphics[width=0.94\textwidth]{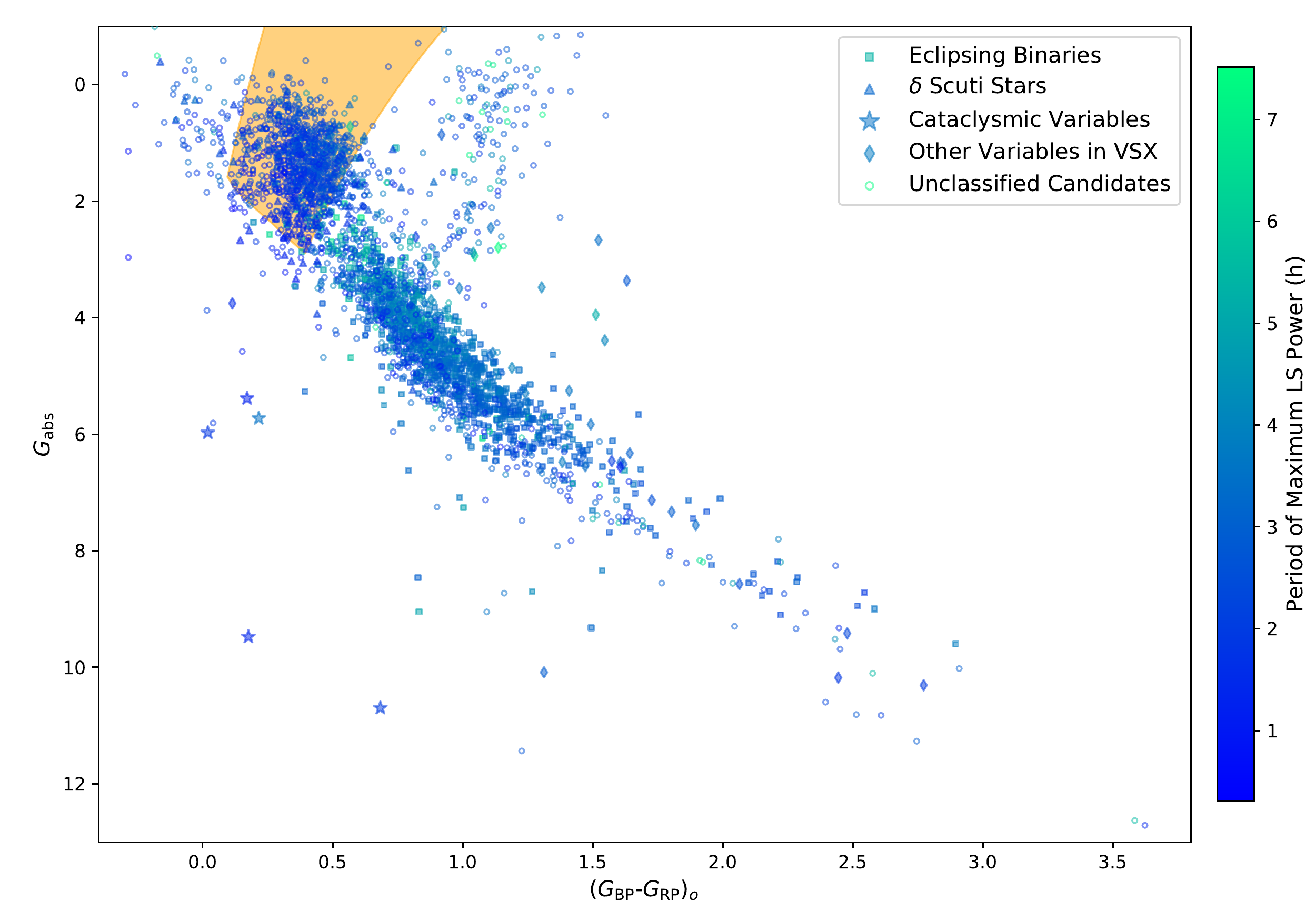}
    \caption{
    \textbf{Distribution of periodic variable star candidates across the HR diagram.}
    { 
    All points represent the periodic variable star candidates selected from the \tmtsdata dataset, among which the eclipsing binaries, $\delta$ Scuti stars, CVs and other variables identified in the International Variable Star Index (VSX)} are highlighted using the symbol shape of square, triangle, star and diamond respectively.
    The open circles represent the periodic variable candidates that are not registered in VSX yet. 
    The color depth of symbols represents the period corresponding to the maximum power in LSP(s).
    The orange area indicates the instability strip.
    Notice that, the period corresponding to eclipsing binaries here is the photometric period rather than the orbital period, since these candidates are not completely classified yet.
     } 
    \label{fig:periodic_variables_hr}
\end{figure*}

The \tmtsdata light curves with periodic variations (typically shorter than 4~hours) are selected by the following criteria: (i) \tmtslocal periodic FAP < 0.001; (ii) no significant Red Noise; (iii) LC quality > 95\%. { As a result, 6626 light curves were selected. 
However, these selection criteria may introduce a large number of light curves with false periodic variations due to that a wide periodicity criteria was applied. In order to improve the true positive rate (TPR), 
we applied two sets of criteria to select the light curves with different ratios of photometric period to observation duration ($P/T$), respectively.
To select light curves that cover at least two complete periods of light variations [i.e., new criterion (iv) $P \leq T/2$], we followed previous selection criteria (i)--(iii). This resulted in 2835 candidate light curves.
For the purpose of collecting light curves with periodic variations slightly longer than a half of the observation duration (i.e., $ T/2 < P \leq 2T/3$), we adopted the following tighter criteria: (i+) \tmtslocal periodic FAP $< 10^{-5}$; (ii) and (iii) are the same as the previous criteria; (iv+) $ T/2 < P \leq 2T/3$ ; (v) the variability index $\epsilon_{\frac{1}{\eta}} \geq 3.0$ (see details in Section~\ref{sec:variability_detection}). With these modified criteria, 988 additional light curves were selected. 

The period distribution of all 3823 candidate light curves (corresponding to 3723 Gaia DR2 sources) is shown in Figure~\ref{fig:periodic_variables_dist}.}
The periods of these selected candidates are distributed from about 20 minutes to 7.5 hours (see Figure~\ref{fig:periodic_variables_dist}).
The number of candidates increases linearly with the photometric period until it peaks at around 3~hours.
For the uninterrupted observations, the detectable period is obviously dependent on the observation duration within a night. Although the observation strategy prevent us from discovering longer-period variable stars, the number of short-period variables found by TMTS is very competitive (see also \citealt{ZTF+DR1+2020,Burdge+etal+2020+systematic,Chen+etal+2020+ZTF_periodic}). For a 5-year survey plan,  TMTS is expected to reveal more than 20,000 periodic variable stars with period shorter than 8 hours.

It is worth noting that, 81 periodic variable star candidates (corresponding to 77 unique sources) have a photometric period below 1~hour,
as the total number of periodic variable stars with periods below 1~hour in the International Variable Star Index (VSX, the version updated on May 31, 2021, \citealt{Watson+etal+2006+VSX}) is only 887.
In the next few years, TMTS will greatly expand the sample of ultra-short periodic variable stars.
The vast majority of these sources are $\delta$ Scuti stars, some blue large-amplitude pulsators (BLAPs, \citealt{Pietrukowicz+etal+2017+blap}) and ultracompact binaries are also likely captured. To distinguish the {BLAPs and} UCBs from the $\delta$ Scuti stars, we need to investigate their absolute magnitudes, colors and spectra.
{
The phase-folded light curves for two shortest-period stars are shown in Figure~\ref{fig:phased_lc_examples}. Their periods (and amplitudes) are 18.4~minutes (0.03~mag) and 18.9~minutes (0.26~mag), respectively.
With the \emph{Gaia} $G_{\rm abs}$ and $G_{\rm BP}-G_{\rm RP}$ color measurements, we inferred that the former is a low-amplitude $\delta$ Scuti star while the latter is a BLAP.
Further photometric and spectroscopic observations are undergoing.
We will study these ultra-short periodic variable stars in separate papers.}

All objects of these candidate light curves were cross-matched with the VSX. 
 As a result, 1603 light curves have a VSX counterpart recorded and almost all these counterparts have a period measurement given by VSX. We plotted the period corresponding to the maximum power in TMTS LSP against the VSX period in Figure~\ref{fig:period_vsx}.
 The VSX periods basically coincide with the TMTS periods or twice the TMTS periods (typically eclipsing binaries) except for a few inconsistent measurements ($\lesssim 2\%$ of all samples). These inconsistent measurements are typically caused by multi-period variable stars or spurious periods in light curves.

All objects corresponding to candidate light curves are also cross-matched with the Gaia DR2 catalogue. Because some light curves match with repeated sources,  all these candidate light curves correspond to \textbf{3723} unique Gaia sources.
Among these stars, \textbf{2987} sources have both a $G_{\rm BP}-G_{\rm RP}$ measurement and a reliable parallax measurement. 
Figure~\ref{fig:periodic_variables_hr} shows the distribution of these sources across the HR diagram. 
Both the absolute magnitudes and the color of these sources were dereddened using the 3D dust map from \cite{Green+etal+2019+3dmap} and the \emph{DUSTMAPS Python} package\footnote{https://github.com/gregreen/dustmaps} \citep{Green+2018+python}. 
After crosschecking with the latest VSX catalogue, we find that more than half of our periodic variable star candidates are newly discovered.
In Figure~\ref{fig:periodic_variables_hr}, we compared these periodic variable star candidates with the $\delta$ Scuti instability strip inferred by using the strip boundaries from \cite{Murphy+etal+2019+ds} and the evolutionary tracks of single stars from the Podova Stellar Evolution Database \citep{Girardi+etal+2000+track}.
As a result, about 940 periodic variable stars were found to locate in the instability strip, 
among which about 800 candidates could be new-discovered $\delta$ Scuti stars.
Notice that, the number is still underestimated, as more than 700 candidates have not good parallax measurement and thus cannot be put in the HR diagram.
This demonstrates the high efficiency of discovering short-period variables by the TMTS.

In comparison, the estimate of the number for eclipsing binaries is more difficult, as the eclipsing binaries distributed on a much wider area of the HR diagram than the $\delta$ Scuti stars.
Due to the special profiles of light curves from eclipsing binaries, they could be identified by the random forest (RF) or neural networks (NNs). By adopting the cyclic-permutation invariant neural networks from \cite{Zhang+Bloom+2021+nn}, there are about 1900 eclipsing binaries in our 3723 periodic variable star candidates, among which about 600 eclipsing binaries are new-discovered (see details in Xi et al. in prep.).
Notice that, the results did not include the eclipsing binaries that were covered by TMTS observations for shorter than 1.5 photometric periods (i.e., 0.75 orbital period). These longer-period eclipsing binaries are expected to be identified by eyes or new algorithm of neural networks.

\begin{figure}
    \includegraphics[width=0.47\textwidth]{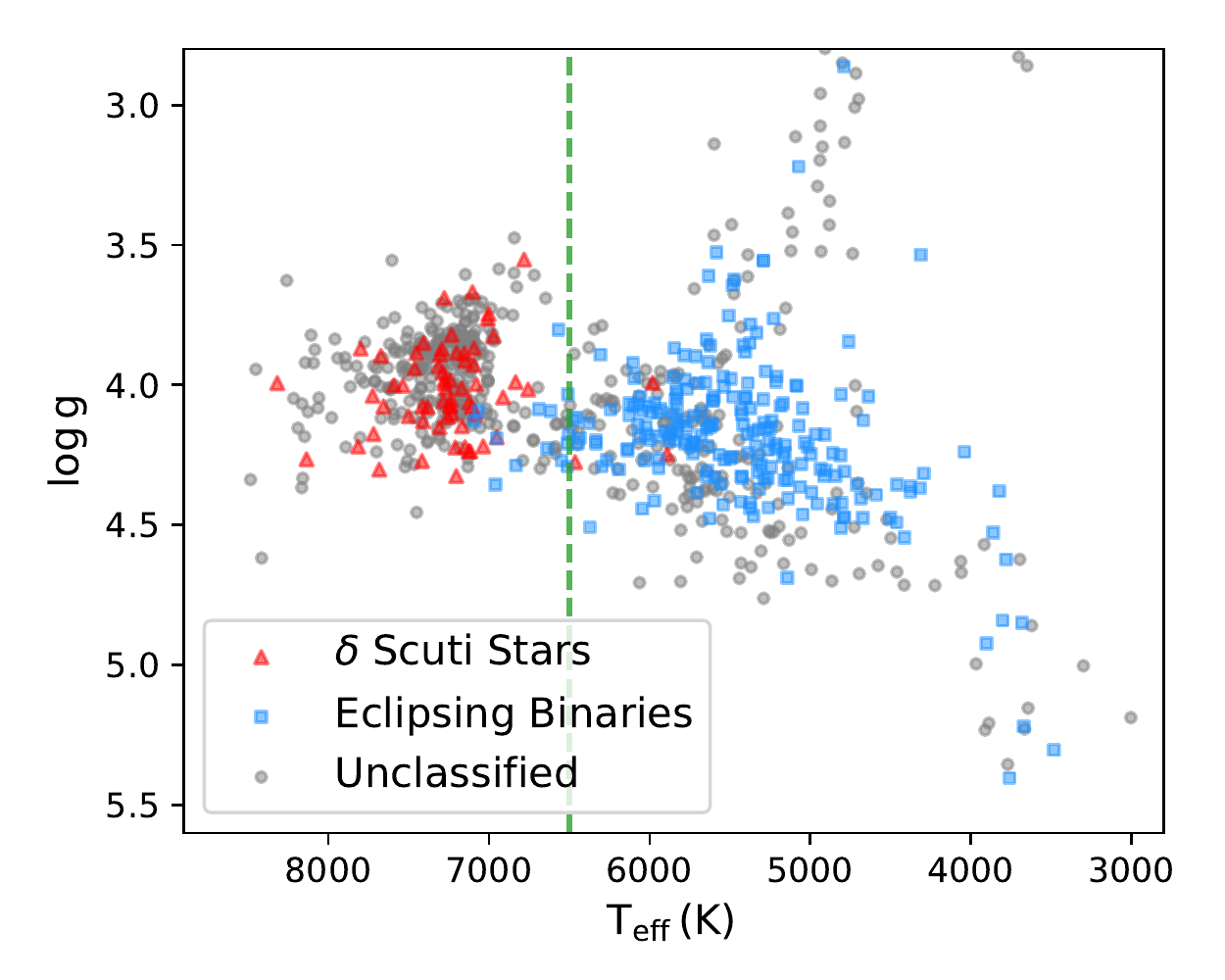}
    \caption{
    \textbf{Lamost surface gravity versus effective temperature for periodic variable candidates from the \tmtsdata dataset.}
    The red triangles and blue squares represent the $\delta$ Scuti stars and eclipsing binaries identified in VSX, respectively. The unclassified periodic variable candidates are labeled as grey circles. The green dashed line indicates the cut-off temperature for selecting $\delta$ Scuti stars (see also \citealt{Murphy+etal+2019+ds}).
     } 
    \label{fig:periodic_variables_logg_teff}
\end{figure}

{

Since about 20\% of periodic variable star candidates have not reliable \emph{Gaia} parallax measurements, the identifications of these candidates rely on additional spectroscopic observations. 
Among 3723 periodic variable star candidates, 1252 sources are observed by the LAMOST and 814 sources have measurements of both surface gravity (log$\,$g) measurement and effective temperature ($\rm T_{eff}$), which means that a part of candidates can be identified using the parameters from the LAMOST spectra.
This is because the two dominant classes of short-period variable sources, $\delta$ Scuti stars and eclipsing binaries,  locate at distinct areas in the $\rm T_{eff}$--log$\,$g diagram (see the Figure~\ref{fig:periodic_variables_logg_teff}).
With the cut-off temperature 6500~K from \cite{Murphy+etal+2019+ds}, 290 new $\delta$ Scuti star candidates are selected from the 502 unidentified periodic variable candidates, and the remained unidentified candidates are very likely eclipsing binaries. 
}

 The LAMOST has started time-domain medium-resolution spectroscopic survey since 2018, and one of the main scientific goals of this survey is to discover quiescent or noninteracting black hole binaries in our Galaxy.
 Since a few binaries harboring an invisible high-mass companion (e.g. $M_{\rm unseen} > 3~M_\odot $) have been proposed as black hole binary candidates \citep{Thompson+etal+2019+Science,Liu+etal+2019+Nature,Gu+etal+2019+bh,Zheng+etal+2019+lamost_bh,Yi+etal+2019+theory_bh,Clavel+etal+2021+bh_test,Gomel+etal+2021+ellipsoidals_III}. 
 With multi-epoch RV measurements from the spectroscopic surveys and periodic provided by the wide-field photometric survey missions like TMTS, the lower limits of mass functions are easily estimated for the observed binaries.
It is worth noting that, the multi-epoch spectra tend to discover short-period binaries,
 since these systems usually have larger Keplerian velocities.
 As introduced by \cite{Yang+etal+2020+EB_lamost}, the most spectroscopic binaries (SBs) identified from the LAMOST have an orbital period shorter than 0.6~day.
  In the next few years, the TMTS is expected to play an important role in providing a large sample with short-period light variations. With the progress of time-domain survey, LAMOST and TMTS observations will give constrains on mass function for a large number of single-lined binaries and thus provide an opportunity for the search of black hole or neutron star binaries.

\subsection{Flare Stars}
\label{sec:flare_stars}

As a subclass of eruptive variable star, flare stars are observed to exhibit flaring behaviour during which the brightness of stars dramatically increase within a few minutes and then decrease for several hours. Since the operation of \emph{Kepler} mission, about 3,400 stars are found to have flares according to the Q1--Q17 (Data Release 25) long-cadence (LC) data of \emph{Kepler} \citep{Yang+Liu+2019+flare_catalog}. However only about 200 flare stars have been identified through its short-cadence (SC) data, since only about 5,000 targets in \emph{Kepler} have SC observations
\citep{Balona+etal+2015+flares,Yang+etal+2018+SC_LC_flares}. Compared to the LC data, the good time resolution of SC data allows further study about the morphology of flares \citep{Balona+etal+2015+flares}.

During the high-cadence observations of the LAMOST sky area, TMTS has also captured a series of flare events. 
Because the flares only cause occasional variations in the light curves, the variability indexes tend to be less sensitive compared with the periodic variations.
Therefore, we widened the criterion of $\epsilon_{\frac{1}{\eta}}$ here when identifying flares.
On the other hand, the flare search is very dependent on the observational condition since two continuous outliers in a light curve are already enough to produce a flare candidate.
It is worth noting that, some spurious ``flares'' could be generated when the objects locate on the bad pixels of detectors. These bad pixels usually distort the point spread function (PSF) of the sources, so the SExtractor Flags of the corresponding measurements are very likely non-zero values. Hence, a very tight criterion is needed for the LC quality. 

\begin{figure}
    \includegraphics[width=0.47\textwidth]{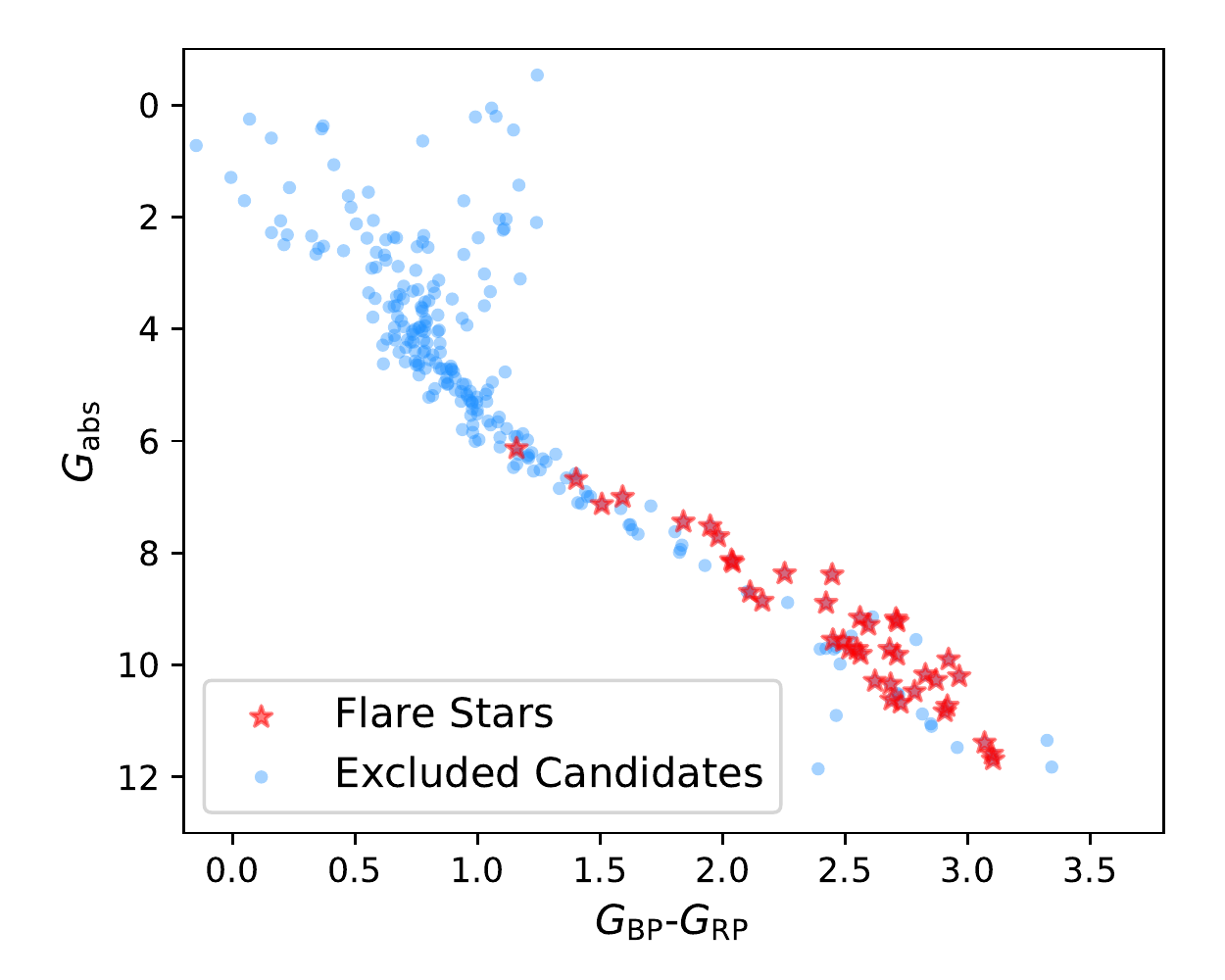}
    \caption{
    \textbf{Distribution of the flare stars and candidates in the HR diagram.}
    The confirmed flare stars are labeled with red star symbol.
     } 
    \label{fig:flar_star_hr}
\end{figure}

In summary, we searched for the flares in the \tmtsdata light curves using the following criteria: (i) \tmtslocal flare FDP $< 10^{-5}$; (ii) $\epsilon_{\frac{1}{\eta}} > 2.0$ (iii) LC quality = 100\% (iv) excluding bad weather observations. 
This finally resulted in 356 flare candidates, among which 42 flares were visually confirmed.
Among them, 39 flares correspond to Gaia sources with both the $G_{\rm BP}-G_{\rm RP}$ color and the reliable parallax measurement.
As shown in Figure~\ref{fig:flar_star_hr}, these flare stars are distributed on the lower right branch of the main sequence in the HR diagram, with an absolute magnitude lying between 6.0~{\rm mag} and 12.0~ {\rm mag}.
Compared to flare stars discovered by other surveys, the 1-min cadence of uninterrupted photometry allow us to further study the shapes and durations (a flare sample is shown in the row \emph{v} of Figure~\ref{fig:lsp_and_phivv} ). 
In a 5-year survey plan of TMTS, the discovery number of flares with  high-quality light curve coverage  is expected to significantly expand the samples of flares with short time resolution.

\section{Summary}

We present the methodology of variables detecting and preliminary scientific results for the first-year high-cadence monitoring of the LAMOST plates with the TMTS.
During the period from January 2020 to December 2020, the TMTS has observed 188 LAMOST plates and generated 4.9~million uninterrupted light curves for over 4.2~million objects with a cadence of about 1 minute.
We have applied the inverse von Neumann ratio, Lomb–Scargle periodogram and \emph{Osten}'s methods to detect variability, periodicity and flares in these light curves, respectively, and then estimated the corresponding significance by their cumulative functions.

A preliminary result of periodicity detection reveals 3723 short-period variable candidates, with a period shorter than 7.5 hours. Hence, TMTS is expected to find  more than 20,000 short-period periodic variables in a five-year observation plan. By plotting the TMTS sources across the HR diagram using the Gaia DR2 parameters, we estimated that at least 600 new eclipsing binaries and 800 new $\delta$ Scuti stars were discovered in 2020. Furthermore, over 40 flares with great temporal resolution were detected during the observations. Further analysis of the periodic variables and flares will be presented in the forthcoming papers.

\section*{Acknowledgements}
We acknowledge the support of the staffs from Xinglong Observatory of NAOC during the installation, commissioning and operation of the TMTS system. This work is supported by the Ma Huateng Foundation, the National Natural Science Foundation of China (NSFC grants 12033003 and 11633002), the National Program on Key Research and Development Project (grant no. 2016YFA0400803). X.W. is also supported by the Scholar Program of Beijing Academy of Science and Technology (BS2020002) and the Strategic Priority Research Program of the Chinese Academy of Sciences, Grant No. XDB23040100.

Guoshoujing Telescope (the Large Sky Area Multi-Object Fiber Spectroscopic Telescope LAMOST) is a National Major Scientific Project built by the Chinese Academy of Sciences. Funding for the project has been provided by the National Development and Reform Commission. LAMOST is operated and managed by the National Astronomical Observatories, Chinese Academy of Sciences.

This work has made use of data from the European Space Agency (ESA) mission
{\it Gaia} (\url{https://www.cosmos.esa.int/gaia}), processed by the {\it Gaia}
Data Processing and Analysis Consortium (DPAC,
\url{https://www.cosmos.esa.int/web/gaia/dpac/consortium}). Funding for the DPAC
has been provided by national institutions, in particular the institutions
participating in the {\it Gaia} Multilateral Agreement.

{
This paper includes data collected with the TESS mission, obtained from the MAST data archive at the Space Telescope Science Institute (STScI). Funding for the TESS mission is provided by the NASA Explorer Program. STScI is operated by the Association of Universities for Research in Astronomy, Inc., under NASA contract NAS 5–26555.

This research has made use of the International Variable Star Index (VSX, \citealt{Watson+etal+2006+VSX}) database, operated at AAVSO, Cambridge, Massachusetts, USA.}
Some of the results in this paper have been derived using the HEALPix \citep{Gorski+etal+2005} package.

\section*{Data Availability}
The data underlying this article are subject to an embargo of about 10 months from the publication date of the article. But the data can still be shared on reasonable request to the corresponding author. The light curves from the first-year survey of TMTS will be publicly available at TMTS Public Data Release 1 (in prep.) in 2022.




\begin{thebibliography}{}
\makeatletter
\relax
\def\mn@urlcharsother{\let\do\@makeother \do\$\do\&\do\#\do\^\do\_\do\%\do\~}
\def\mn@doi{\begingroup\mn@urlcharsother \@ifnextchar [ {\mn@doi@}
  {\mn@doi@[]}}
\def\mn@doi@[#1]#2{\def\@tempa{#1}\ifx\@tempa\@empty \href
  {http://dx.doi.org/#2} {doi:#2}\else \href {http://dx.doi.org/#2} {#1}\fi
  \endgroup}
\def\mn@eprint#1#2{\mn@eprint@#1:#2::\@nil}
\def\mn@eprint@arXiv#1{\href {http://arxiv.org/abs/#1} {{\tt arXiv:#1}}}
\def\mn@eprint@dblp#1{\href {http://dblp.uni-trier.de/rec/bibtex/#1.xml}
  {dblp:#1}}
\def\mn@eprint@#1:#2:#3:#4\@nil{\def\@tempa {#1}\def\@tempb {#2}\def\@tempc
  {#3}\ifx \@tempc \@empty \let \@tempc \@tempb \let \@tempb \@tempa \fi \ifx
  \@tempb \@empty \def\@tempb {arXiv}\fi \@ifundefined
  {mn@eprint@\@tempb}{\@tempb:\@tempc}{\expandafter \expandafter \csname
  mn@eprint@\@tempb\endcsname \expandafter{\@tempc}}}

\bibitem[\protect\citeauthoryear{{Abramowitz} \& {Stegun}}{{Abramowitz} \&
  {Stegun}}{1972}]{Abramowitz+Stegun+1972+mathematical+function}
{Abramowitz} M.,  {Stegun} I.~A.,  1972, {Handbook of Mathematical Functions}

\bibitem[\protect\citeauthoryear{{Aigrain}, {Parviainen}, {Roberts}, {Reece}
  \& {Evans}}{{Aigrain} et~al.}{2017}]{Aigrain+etal+2017+sysrem}
{Aigrain} S.,  {Parviainen} H.,  {Roberts} S.,  {Reece} S.,   {Evans} T.,
  2017, \mn@doi [\mnras] {10.1093/mnras/stx1422}, \href
  {https://ui.adsabs.harvard.edu/abs/2017MNRAS.471..759A} {471, 759}

\bibitem[\protect\citeauthoryear{{Alard} \& {Lupton}}{{Alard} \&
  {Lupton}}{1998}]{Alard+Lupton+1998+DIA}
{Alard} C.,  {Lupton} R.~H.,  1998, \mn@doi [\apj] {10.1086/305984}, \href
  {https://ui.adsabs.harvard.edu/abs/1998ApJ...503..325A} {503, 325}

\bibitem[\protect\citeauthoryear{{Bahramian} et~al.,}{{Bahramian}
  et~al.}{2017}]{Bahramian+etal+2017+UCBHXB}
{Bahramian} A.,  et~al., 2017, \mn@doi [\mnras] {10.1093/mnras/stx166}, \href
  {https://ui.adsabs.harvard.edu/abs/2017MNRAS.467.2199B} {467, 2199}

\bibitem[\protect\citeauthoryear{{Balona}}{{Balona}}{2015}]{Balona+etal+2015+flares}
{Balona} L.~A.,  2015, \mn@doi [\mnras] {10.1093/mnras/stu2651}, \href
  {https://ui.adsabs.harvard.edu/abs/2015MNRAS.447.2714B} {447, 2714}

\bibitem[\protect\citeauthoryear{{Baluev}}{{Baluev}}{2008}]{Baluev+2008+FAP_corrected}
{Baluev} R.~V.,  2008, \mn@doi [\mnras] {10.1111/j.1365-2966.2008.12689.x},
  \href {https://ui.adsabs.harvard.edu/abs/2008MNRAS.385.1279B} {385, 1279}

\bibitem[\protect\citeauthoryear{{Becker} et~al.,}{{Becker}
  et~al.}{2004}]{DLS+2004_short}
{Becker} A.~C.,  et~al., 2004, \mn@doi [\apj] {10.1086/421994}, \href
  {https://ui.adsabs.harvard.edu/abs/2004ApJ...611..418B} {611, 418}

\bibitem[\protect\citeauthoryear{{Bellm} et~al.,}{{Bellm}
  et~al.}{2019}]{ZTF+2019+first}
{Bellm} E.~C.,  et~al., 2019, \mn@doi [\pasp] {10.1088/1538-3873/aaecbe}, \href
  {https://ui.adsabs.harvard.edu/abs/2019PASP..131a8002B} {131, 018002}

\bibitem[\protect\citeauthoryear{{Bertin}}{{Bertin}}{2006}]{Bertin+2006+SCAMP}
{Bertin} E.,  2006, in {Gabriel} C.,  {Arviset} C.,  {Ponz} D.,   {Enrique} S.,
   eds,  Astronomical Society of the Pacific Conference Series Vol. 351,
  Astronomical Data Analysis Software and Systems XV. p.~112

\bibitem[\protect\citeauthoryear{{Bertin} \& {Arnouts}}{{Bertin} \&
  {Arnouts}}{1996}]{Bertin+Arnouts+1996+SExtractor}
{Bertin} E.,  {Arnouts} S.,  1996, \mn@doi [\aaps] {10.1051/aas:1996164}, \href
  {https://ui.adsabs.harvard.edu/abs/1996A&AS..117..393B} {117, 393}

\bibitem[\protect\citeauthoryear{{Bertin}, {Mellier}, {Radovich}, {Missonnier},
  {Didelon}  \& {Morin}}{{Bertin} et~al.}{2002}]{Bertin+etal+2002+TERAPIX}
{Bertin} E.,  {Mellier} Y.,  {Radovich} M.,  {Missonnier} G.,  {Didelon} P.,
  {Morin} B.,  2002, in {Bohlender} D.~A.,  {Durand} D.,   {Handley} T.~H.,
  eds,  Astronomical Society of the Pacific Conference Series Vol. 281,
  Astronomical Data Analysis Software and Systems XI. p.~228

\bibitem[\protect\citeauthoryear{{Borucki} et~al.,}{{Borucki}
  et~al.}{2010}]{Kepler+2010+first}
{Borucki} W.~J.,  et~al., 2010, \mn@doi [Science] {10.1126/science.1185402},
  \href {https://ui.adsabs.harvard.edu/abs/2010Sci...327..977B} {327, 977}

\bibitem[\protect\citeauthoryear{{Burdge} et~al.,}{{Burdge}
  et~al.}{2019}]{Burdge+etal+2019+Nature}
{Burdge} K.~B.,  et~al., 2019, \mn@doi [\nat] {10.1038/s41586-019-1403-0},
  \href {https://ui.adsabs.harvard.edu/abs/2019Natur.571..528B} {571, 528}

\bibitem[\protect\citeauthoryear{{Burdge} et~al.,}{{Burdge}
  et~al.}{2020a}]{Burdge+etal+2020+systematic}
{Burdge} K.~B.,  et~al., 2020a, arXiv e-prints, \href
  {https://ui.adsabs.harvard.edu/abs/2020arXiv200902567B} {p. arXiv:2009.02567}

\bibitem[\protect\citeauthoryear{{Burdge} et~al.,}{{Burdge}
  et~al.}{2020b}]{Burdge+etal+2020+8.8min}
{Burdge} K.~B.,  et~al., 2020b, arXiv e-prints, \href
  {https://ui.adsabs.harvard.edu/abs/2020arXiv201003555B} {p. arXiv:2010.03555}

\bibitem[\protect\citeauthoryear{{Chatterjee}, {Nugent}, {Brady}, {Cannella},
  {Kaplan}  \& {Kasliwal}}{{Chatterjee} et~al.}{2019}]{iPTF+2019+detectability}
{Chatterjee} D.,  {Nugent} P.~E.,  {Brady} P.~R.,  {Cannella} C.,  {Kaplan}
  D.~L.,   {Kasliwal} M.~M.,  2019, \mn@doi [\apj] {10.3847/1538-4357/ab2b9c},
  \href {https://ui.adsabs.harvard.edu/abs/2019ApJ...881..128C} {881, 128}

\bibitem[\protect\citeauthoryear{{Chen}, {Wang}, {Deng}, {de Grijs}, {Yang}  \&
  {Tian}}{{Chen} et~al.}{2020a}]{Chen+etal+2020+ZTF_periodic}
{Chen} X.,  {Wang} S.,  {Deng} L.,  {de Grijs} R.,  {Yang} M.,   {Tian} H.,
  2020a, \mn@doi [\apjs] {10.3847/1538-4365/ab9cae}, \href
  {https://ui.adsabs.harvard.edu/abs/2020ApJS..249...18C} {249, 18}

\bibitem[\protect\citeauthoryear{{Chen}, {Liu}  \& {Wang}}{{Chen}
  et~al.}{2020b}]{Chen+etal+2020}
{Chen} W.-C.,  {Liu} D.-D.,   {Wang} B.,  2020b, \mn@doi [\apjl]
  {10.3847/2041-8213/abae66}, \href
  {https://ui.adsabs.harvard.edu/abs/2020ApJ...900L...8C} {900, L8}

\bibitem[\protect\citeauthoryear{{Clavel}, {Dubus}, {Casares}  \&
  {Babusiaux}}{{Clavel} et~al.}{2021}]{Clavel+etal+2021+bh_test}
{Clavel} M.,  {Dubus} G.,  {Casares} J.,   {Babusiaux} C.,  2021, \mn@doi
  [\aap] {10.1051/0004-6361/202039317}, \href
  {https://ui.adsabs.harvard.edu/abs/2021A&A...645A..72C} {645, A72}

\bibitem[\protect\citeauthoryear{{Coughlin} et~al.,}{{Coughlin}
  et~al.}{2020}]{Coughlin+etal+2020+ZTFprojectionII}
{Coughlin} M.~W.,  et~al., 2020, arXiv e-prints, \href
  {https://ui.adsabs.harvard.edu/abs/2020arXiv200914071C} {p. arXiv:2009.14071}

\bibitem[\protect\citeauthoryear{{Cui} et~al.,}{{Cui}
  et~al.}{2012}]{Cui+etal+2012+LAMOST}
{Cui} X.-Q.,  et~al., 2012, \mn@doi [Research in Astronomy and Astrophysics]
  {10.1088/1674-4527/12/9/003}, \href
  {https://ui.adsabs.harvard.edu/abs/2012RAA....12.1197C} {12, 1197}

\bibitem[\protect\citeauthoryear{{Downes}, {Webbink}, {Shara}, {Ritter}, {Kolb}
   \& {Duerbeck}}{{Downes} et~al.}{2001}]{Downes+etal+2001+CV}
{Downes} R.~A.,  {Webbink} R.~F.,  {Shara} M.~M.,  {Ritter} H.,  {Kolb} U.,
  {Duerbeck} H.~W.,  2001, \mn@doi [\pasp] {10.1086/320802}, \href
  {https://ui.adsabs.harvard.edu/abs/2001PASP..113..764D} {113, 764}

\bibitem[\protect\citeauthoryear{{Drake} et~al.,}{{Drake}
  et~al.}{2009}]{Catalina+2009+first}
{Drake} A.~J.,  et~al., 2009, \mn@doi [\apj] {10.1088/0004-637X/696/1/870},
  \href {https://ui.adsabs.harvard.edu/abs/2009ApJ...696..870D} {696, 870}

\bibitem[\protect\citeauthoryear{{Drake} et~al.,}{{Drake}
  et~al.}{2013}]{Drake+etal+2013}
{Drake} A.~J.,  et~al., 2013, \mn@doi [\apj] {10.1088/0004-637X/763/1/32},
  \href {https://ui.adsabs.harvard.edu/abs/2013ApJ...763...32D} {763, 32}

\bibitem[\protect\citeauthoryear{{Drake} et~al.,}{{Drake}
  et~al.}{2014a}]{Drake+etal+2014+catalina_period}
{Drake} A.~J.,  et~al., 2014a, \mn@doi [\apjs] {10.1088/0067-0049/213/1/9},
  \href {https://ui.adsabs.harvard.edu/abs/2014ApJS..213....9D} {213, 9}

\bibitem[\protect\citeauthoryear{{Drake} et~al.,}{{Drake}
  et~al.}{2014b}]{Catelina+2014+CV}
{Drake} A.~J.,  et~al., 2014b, \mn@doi [\mnras] {10.1093/mnras/stu639}, \href
  {https://ui.adsabs.harvard.edu/abs/2014MNRAS.441.1186D} {441, 1186}

\bibitem[\protect\citeauthoryear{{Drake} et~al.,}{{Drake}
  et~al.}{2014c}]{Catalina+2014+ultracompact}
{Drake} A.~J.,  et~al., 2014c, \mn@doi [\apj] {10.1088/0004-637X/790/2/157},
  \href {https://ui.adsabs.harvard.edu/abs/2014ApJ...790..157D} {790, 157}

\bibitem[\protect\citeauthoryear{{Drake} et~al.,}{{Drake}
  et~al.}{2017}]{Drake+etal+2017+catalina_period}
{Drake} A.~J.,  et~al., 2017, \mn@doi [\mnras] {10.1093/mnras/stx1085}, \href
  {https://ui.adsabs.harvard.edu/abs/2017MNRAS.469.3688D} {469, 3688}

\bibitem[\protect\citeauthoryear{{Gaia Collaboration} et~al.,}{{Gaia
  Collaboration} et~al.}{2016}]{Gaia_Collaboration+2016+performance}
{Gaia Collaboration} et~al., 2016, \mn@doi [\aap]
  {10.1051/0004-6361/201629272}, \href
  {https://ui.adsabs.harvard.edu/abs/2016A&A...595A...1G} {595, A1}

\bibitem[\protect\citeauthoryear{{Gaia Collaboration} et~al.,}{{Gaia
  Collaboration} et~al.}{2018}]{Gaia_collaboration+2018+data}
{Gaia Collaboration} et~al., 2018, \mn@doi [\aap]
  {10.1051/0004-6361/201833051}, \href
  {https://ui.adsabs.harvard.edu/abs/2018A&A...616A...1G} {616, A1}

\bibitem[\protect\citeauthoryear{{Geier}}{{Geier}}{2020}]{Geier+2020+subdwarf}
{Geier} S.,  2020, \mn@doi [\aap] {10.1051/0004-6361/202037526}, \href
  {https://ui.adsabs.harvard.edu/abs/2020A&A...635A.193G} {635, A193}

\bibitem[\protect\citeauthoryear{{Geier}, {Raddi}, {Gentile Fusillo}  \&
  {Marsh}}{{Geier} et~al.}{2019}]{Geier+etal+2019+hot_subdwarf}
{Geier} S.,  {Raddi} R.,  {Gentile Fusillo} N.~P.,   {Marsh} T.~R.,  2019,
  \mn@doi [\aap] {10.1051/0004-6361/201834236}, \href
  {https://ui.adsabs.harvard.edu/abs/2019A&A...621A..38G} {621, A38}

\bibitem[\protect\citeauthoryear{{Gentile Fusillo} et~al.,}{{Gentile Fusillo}
  et~al.}{2019}]{Fusillo+etal+2019+gaia_wd}
{Gentile Fusillo} N.~P.,  et~al., 2019, \mn@doi [\mnras]
  {10.1093/mnras/sty3016}, \href
  {https://ui.adsabs.harvard.edu/abs/2019MNRAS.482.4570G} {482, 4570}

\bibitem[\protect\citeauthoryear{{Girardi}, {Bressan}, {Bertelli}  \&
  {Chiosi}}{{Girardi} et~al.}{2000}]{Girardi+etal+2000+track}
{Girardi} L.,  {Bressan} A.,  {Bertelli} G.,   {Chiosi} C.,  2000, \mn@doi
  [\aaps] {10.1051/aas:2000126}, \href
  {https://ui.adsabs.harvard.edu/abs/2000A&AS..141..371G} {141, 371}

\bibitem[\protect\citeauthoryear{{Gomel}, {Faigler}  \& {Mazeh}}{{Gomel}
  et~al.}{2021a}]{Gomel+etal+2021+ellipsoidals_I}
{Gomel} R.,  {Faigler} S.,   {Mazeh} T.,  2021a, \mn@doi [\mnras]
  {10.1093/mnras/staa3305}, \href
  {https://ui.adsabs.harvard.edu/abs/2021MNRAS.501.2822G} {501, 2822}

\bibitem[\protect\citeauthoryear{{Gomel}, {Faigler}  \& {Mazeh}}{{Gomel}
  et~al.}{2021b}]{Gomel+etal+2021+ellipsoidals_II}
{Gomel} R.,  {Faigler} S.,   {Mazeh} T.,  2021b, \mn@doi [\mnras]
  {10.1093/mnras/stab1047}, \href
  {https://ui.adsabs.harvard.edu/abs/2021MNRAS.504.2115G} {504, 2115}

\bibitem[\protect\citeauthoryear{{Gomel}, {Faigler}, {Mazeh}  \&
  {Pawlak}}{{Gomel} et~al.}{2021c}]{Gomel+etal+2021+ellipsoidals_III}
{Gomel} R.,  {Faigler} S.,  {Mazeh} T.,   {Pawlak} M.,  2021c, \mn@doi [\mnras]
  {10.1093/mnras/stab1235}, \href
  {https://ui.adsabs.harvard.edu/abs/2021MNRAS.504.5907G} {504, 5907}

\bibitem[\protect\citeauthoryear{{G{\'o}rski}, {Hivon}, {Banday}, {Wandelt},
  {Hansen}, {Reinecke}  \& {Bartelmann}}{{G{\'o}rski}
  et~al.}{2005}]{Gorski+etal+2005}
{G{\'o}rski} K.~M.,  {Hivon} E.,  {Banday} A.~J.,  {Wandelt} B.~D.,  {Hansen}
  F.~K.,  {Reinecke} M.,   {Bartelmann} M.,  2005, \mn@doi [\apj]
  {10.1086/427976}, \href
  {https://ui.adsabs.harvard.edu/abs/2005ApJ...622..759G} {622, 759}

\bibitem[\protect\citeauthoryear{{Green}}{{Green}}{2018}]{Green+2018+python}
{Green} G.~M.,  2018, \mn@doi [The Journal of Open Source Software]
  {10.21105/joss.00695}, \href
  {https://ui.adsabs.harvard.edu/abs/2018JOSS....3..695G} {3, 695}

\bibitem[\protect\citeauthoryear{{Green}, {Schlafly}, {Zucker}, {Speagle}  \&
  {Finkbeiner}}{{Green} et~al.}{2019}]{Green+etal+2019+3dmap}
{Green} G.~M.,  {Schlafly} E.,  {Zucker} C.,  {Speagle} J.~S.,   {Finkbeiner}
  D.,  2019, \mn@doi [\apj] {10.3847/1538-4357/ab5362}, \href
  {https://ui.adsabs.harvard.edu/abs/2019ApJ...887...93G} {887, 93}

\bibitem[\protect\citeauthoryear{{Groot} et~al.,}{{Groot}
  et~al.}{2003}]{FSVS+2003+reduction}
{Groot} P.~J.,  et~al., 2003, \mn@doi [\mnras]
  {10.1046/j.1365-8711.2003.06182.x}, \href
  {https://ui.adsabs.harvard.edu/abs/2003MNRAS.339..427G} {339, 427}

\bibitem[\protect\citeauthoryear{{Gu} et~al.,}{{Gu}
  et~al.}{2019}]{Gu+etal+2019+bh}
{Gu} W.-M.,  et~al., 2019, \mn@doi [\apjl] {10.3847/2041-8213/ab04f0}, \href
  {https://ui.adsabs.harvard.edu/abs/2019ApJ...872L..20G} {872, L20}

\bibitem[\protect\citeauthoryear{{Ho} et~al.,}{{Ho}
  et~al.}{2018}]{Ho+etal+2018+iptf}
{Ho} A. Y.~Q.,  et~al., 2018, \mn@doi [\apjl] {10.3847/2041-8213/aaaa62}, \href
  {https://ui.adsabs.harvard.edu/abs/2018ApJ...854L..13H} {854, L13}

\bibitem[\protect\citeauthoryear{{Huang}, {Li}, {Wang}, {Shang}, {Zhang}, {Hu},
  {Qiu}  \& {Jiang}}{{Huang} et~al.}{2012}]{Huang+etal+2012+TNT}
{Huang} F.,  {Li} J.-Z.,  {Wang} X.-F.,  {Shang} R.-C.,  {Zhang} T.-M.,  {Hu}
  J.-Y.,  {Qiu} Y.-L.,   {Jiang} X.-J.,  2012, \mn@doi [Research in Astronomy
  and Astrophysics] {10.1088/1674-4527/12/11/012}, \href
  {https://ui.adsabs.harvard.edu/abs/2012RAA....12.1585H} {12, 1585}

\bibitem[\protect\citeauthoryear{{Ivezi{\'c}}, {Connelly}, {VanderPlas}  \&
  {Gray}}{{Ivezi{\'c}} et~al.}{2014}]{Ivezi+2014+book}
{Ivezi{\'c}} {\v{Z}}.,  {Connelly} A.~J.,  {VanderPlas} J.~T.,   {Gray} A.,
  2014, {Statistics, Data Mining, and Machine Learning in Astronomy}

\bibitem[\protect\citeauthoryear{{Jim{\'e}nez-Esteban}, {Torres},
  {Rebassa-Mansergas}, {Skorobogatov}, {Solano}, {Cantero}  \&
  {Rodrigo}}{{Jim{\'e}nez-Esteban} et~al.}{2018}]{Esteban+etal+2018+gaia_wd}
{Jim{\'e}nez-Esteban} F.~M.,  {Torres} S.,  {Rebassa-Mansergas} A.,
  {Skorobogatov} G.,  {Solano} E.,  {Cantero} C.,   {Rodrigo} C.,  2018,
  \mn@doi [\mnras] {10.1093/mnras/sty2120}, \href
  {https://ui.adsabs.harvard.edu/abs/2018MNRAS.480.4505J} {480, 4505}

\bibitem[\protect\citeauthoryear{{Kim} \& {Bailer-Jones}}{{Kim} \&
  {Bailer-Jones}}{2016}]{Kim+etal+2016+UPSILON}
{Kim} D.-W.,  {Bailer-Jones} C. A.~L.,  2016, \mn@doi [\aap]
  {10.1051/0004-6361/201527188}, \href
  {https://ui.adsabs.harvard.edu/abs/2016A&A...587A..18K} {587, A18}

\bibitem[\protect\citeauthoryear{{Kim}, {L{\'e}pine}  \& {Medan}}{{Kim}
  et~al.}{2020}]{Kim+etal+2020+wd}
{Kim} B.,  {L{\'e}pine} S.,   {Medan} I.,  2020, \mn@doi [\apj]
  {10.3847/1538-4357/aba523}, \href
  {https://ui.adsabs.harvard.edu/abs/2020ApJ...899...83K} {899, 83}

\bibitem[\protect\citeauthoryear{{Knevitt}, {Wynn}, {Vaughan}  \&
  {Watson}}{{Knevitt} et~al.}{2014}]{Knevitt+etal+2014}
{Knevitt} G.,  {Wynn} G.~A.,  {Vaughan} S.,   {Watson} M.~G.,  2014, \mn@doi
  [\mnras] {10.1093/mnras/stt2008}, \href
  {https://ui.adsabs.harvard.edu/abs/2014MNRAS.437.3087K} {437, 3087}

\bibitem[\protect\citeauthoryear{{Koch} et~al.,}{{Koch}
  et~al.}{2010}]{Koch+etal+2010+Kepler_intrument}
{Koch} D.~G.,  et~al., 2010, \mn@doi [\apjl] {10.1088/2041-8205/713/2/L79},
  \href {https://ui.adsabs.harvard.edu/abs/2010ApJ...713L..79K} {713, L79}

\bibitem[\protect\citeauthoryear{{Kowalski}, {Hawley}, {Hilton}, {Becker},
  {West}, {Bochanski}  \& {Sesar}}{{Kowalski}
  et~al.}{2009}]{Kowalski+etal+2009+FDR_flare}
{Kowalski} A.~F.,  {Hawley} S.~L.,  {Hilton} E.~J.,  {Becker} A.~C.,  {West}
  A.~A.,  {Bochanski} J.~J.,   {Sesar} B.,  2009, \mn@doi [\aj]
  {10.1088/0004-6256/138/2/633}, \href
  {https://ui.adsabs.harvard.edu/abs/2009AJ....138..633K} {138, 633}

\bibitem[\protect\citeauthoryear{{Kulkarni} \& {Rau}}{{Kulkarni} \&
  {Rau}}{2006}]{Kulkarni+FOTs+DLS+2006}
{Kulkarni} S.~R.,  {Rau} A.,  2006, \mn@doi [\apjl] {10.1086/505423}, \href
  {https://ui.adsabs.harvard.edu/abs/2006ApJ...644L..63K} {644, L63}

\bibitem[\protect\citeauthoryear{{Kupfer} et~al.,}{{Kupfer}
  et~al.}{2021}]{Kupfer+etal+2021}
{Kupfer} T.,  et~al., 2021, \mn@doi [\mnras] {10.1093/mnras/stab1344}, \href
  {https://ui.adsabs.harvard.edu/abs/2021MNRAS.tmp.1321K} {}

\bibitem[\protect\citeauthoryear{{Law} et~al.,}{{Law}
  et~al.}{2009}]{PTF+2009+performance}
{Law} N.~M.,  et~al., 2009, \mn@doi [\pasp] {10.1086/648598}, \href
  {https://ui.adsabs.harvard.edu/abs/2009PASP..121.1395L} {121, 1395}

\bibitem[\protect\citeauthoryear{{Liu} et~al.,}{{Liu}
  et~al.}{2019}]{Liu+etal+2019+Nature}
{Liu} J.,  et~al., 2019, \mn@doi [\nat] {10.1038/s41586-019-1766-2}, \href
  {https://ui.adsabs.harvard.edu/abs/2019Natur.575..618L} {575, 618}

\bibitem[\protect\citeauthoryear{{Liu} et~al.,}{{Liu}
  et~al.}{2020}]{Liu+etal+2020}
{Liu} C.,  et~al., 2020, arXiv e-prints, \href
  {https://ui.adsabs.harvard.edu/abs/2020arXiv200507210L} {p. arXiv:2005.07210}

\bibitem[\protect\citeauthoryear{{Lomb}}{{Lomb}}{1976}]{Lomb+1976}
{Lomb} N.~R.,  1976, \mn@doi [\apss] {10.1007/BF00648343}, \href
  {https://ui.adsabs.harvard.edu/abs/1976Ap&SS..39..447L} {39, 447}

\bibitem[\protect\citeauthoryear{{Macfarlane}, {Toma}, {Ramsay}, {Groot},
  {Woudt}, {Drew}, {Barentsen}  \& {Eisl{\"o}ffel}}{{Macfarlane}
  et~al.}{2015}]{OmegaWhite+2015+survey}
{Macfarlane} S.~A.,  {Toma} R.,  {Ramsay} G.,  {Groot} P.~J.,  {Woudt} P.~A.,
  {Drew} J.~E.,  {Barentsen} G.,   {Eisl{\"o}ffel} J.,  2015, \mn@doi [\mnras]
  {10.1093/mnras/stv1989}, \href
  {https://ui.adsabs.harvard.edu/abs/2015MNRAS.454..507M} {454, 507}

\bibitem[\protect\citeauthoryear{{Marsh}, {Prince}, {Mahabal}, {Bellm}, {Drake}
   \& {Djorgovski}}{{Marsh} et~al.}{2017}]{Marsh+etal+2017+eb}
{Marsh} F.~M.,  {Prince} T.~A.,  {Mahabal} A.~A.,  {Bellm} E.~C.,  {Drake}
  A.~J.,   {Djorgovski} S.~G.,  2017, \mn@doi [\mnras] {10.1093/mnras/stw2110},
  \href {https://ui.adsabs.harvard.edu/abs/2017MNRAS.465.4678M} {465, 4678}

\bibitem[\protect\citeauthoryear{{Mart{\'\i}nez-Palomera}
  et~al.,}{{Mart{\'\i}nez-Palomera} et~al.}{2018}]{HiTS+2018}
{Mart{\'\i}nez-Palomera} J.,  et~al., 2018, \mn@doi [\aj]
  {10.3847/1538-3881/aadfd8}, \href
  {https://ui.adsabs.harvard.edu/abs/2018AJ....156..186M} {156, 186}

\bibitem[\protect\citeauthoryear{{Masci} et~al.,}{{Masci}
  et~al.}{2019}]{ZTF+2019+products}
{Masci} F.~J.,  et~al., 2019, \mn@doi [\pasp] {10.1088/1538-3873/aae8ac}, \href
  {https://ui.adsabs.harvard.edu/abs/2019PASP..131a8003M} {131, 018003}

\bibitem[\protect\citeauthoryear{{Menou}, {Narayan}  \& {Lasota}}{{Menou}
  et~al.}{1999}]{Menou+etal+1999}
{Menou} K.,  {Narayan} R.,   {Lasota} J.-P.,  1999, \mn@doi [\apj]
  {10.1086/306878}, \href
  {https://ui.adsabs.harvard.edu/abs/1999ApJ...513..811M} {513, 811}

\bibitem[\protect\citeauthoryear{{Miller} et~al.,}{{Miller}
  et~al.}{2001}]{Miller+etal+2001+FDR}
{Miller} C.~J.,  et~al., 2001, \mn@doi [\aj] {10.1086/324109}, \href
  {https://ui.adsabs.harvard.edu/abs/2001AJ....122.3492M} {122, 3492}

\bibitem[\protect\citeauthoryear{{Murphy}, {Hey}, {Van Reeth}  \&
  {Bedding}}{{Murphy} et~al.}{2019}]{Murphy+etal+2019+ds}
{Murphy} S.~J.,  {Hey} D.,  {Van Reeth} T.,   {Bedding} T.~R.,  2019, \mn@doi
  [\mnras] {10.1093/mnras/stz590}, \href
  {https://ui.adsabs.harvard.edu/abs/2019MNRAS.485.2380M} {485, 2380}

\bibitem[\protect\citeauthoryear{{Nidever} et~al.,}{{Nidever}
  et~al.}{2021}]{Nidever+etal+2021+survey}
{Nidever} D.~L.,  et~al., 2021, \mn@doi [\aj] {10.3847/1538-3881/abd6e1}, \href
  {https://ui.adsabs.harvard.edu/abs/2021AJ....161..192N} {161, 192}

\bibitem[\protect\citeauthoryear{{Ofek}, {Soumagnac}, {Nir}, {Gal-Yam},
  {Nugent}, {Masci}  \& {Kulkarni}}{{Ofek} et~al.}{2020}]{ZTF+DR1+2020}
{Ofek} E.~O.,  {Soumagnac} M.,  {Nir} G.,  {Gal-Yam} A.,  {Nugent} P.,  {Masci}
  F.,   {Kulkarni} S.~R.,  2020, \mn@doi [\mnras] {10.1093/mnras/staa2814},
  \href {https://ui.adsabs.harvard.edu/abs/2020MNRAS.499.5782O} {499, 5782}

\bibitem[\protect\citeauthoryear{{Ofir} et~al.,}{{Ofir}
  et~al.}{2010}]{Ofir+etal+2010+corot}
{Ofir} A.,  et~al., 2010, \mn@doi [\mnras] {10.1111/j.1745-3933.2010.00843.x},
  \href {https://ui.adsabs.harvard.edu/abs/2010MNRAS.404L..99O} {404, L99}

\bibitem[\protect\citeauthoryear{{Osten}, {Kowalski}, {Sahu}  \&
  {Hawley}}{{Osten} et~al.}{2012}]{Osten+etal+2012+flare_archive}
{Osten} R.~A.,  {Kowalski} A.,  {Sahu} K.,   {Hawley} S.~L.,  2012, \mn@doi
  [\apj] {10.1088/0004-637X/754/1/4}, \href
  {https://ui.adsabs.harvard.edu/abs/2012ApJ...754....4O} {754, 4}

\bibitem[\protect\citeauthoryear{{P{\'a}l}}{{P{\'a}l}}{2012}]{Pal+2012+FITSH}
{P{\'a}l} A.,  2012, \mn@doi [\mnras] {10.1111/j.1365-2966.2011.19813.x}, \href
  {https://ui.adsabs.harvard.edu/abs/2012MNRAS.421.1825P} {421, 1825}

\bibitem[\protect\citeauthoryear{{Pala}, {Ederoclite}, {G{\"a}nsicke}, {Gentile
  Fusillo}, {Abril}, {Raddi}, {V{\'a}zquez Rami{\'o}}  \&
  {Rebassa-Mansergas}}{{Pala} et~al.}{2020a}]{CHiCaS+2020+survey}
{Pala} A.~F.,  {Ederoclite} A.,  {G{\"a}nsicke} B.~T.,  {Gentile Fusillo}
  N.~P.,  {Abril} J.,  {Raddi} R.,  {V{\'a}zquez Rami{\'o}} H.,
  {Rebassa-Mansergas} A.,  2020a, \mn@doi [Advances in Space Research]
  {10.1016/j.asr.2020.05.033}, \href
  {https://ui.adsabs.harvard.edu/abs/2020AdSpR..66.1235P} {66, 1235}

\bibitem[\protect\citeauthoryear{{Pala} et~al.,}{{Pala}
  et~al.}{2020b}]{Pala+etal+2020+gaia_CV}
{Pala} A.~F.,  et~al., 2020b, \mn@doi [\mnras] {10.1093/mnras/staa764}, \href
  {https://ui.adsabs.harvard.edu/abs/2020MNRAS.494.3799P} {494, 3799}

\bibitem[\protect\citeauthoryear{{Paudel}, {Gizis}, {Mullan}, {Schmidt},
  {Burgasser}, {Williams}  \& {Berger}}{{Paudel}
  et~al.}{2018}]{Paudel+etal+2018+K2_flare}
{Paudel} R.~R.,  {Gizis} J.~E.,  {Mullan} D.~J.,  {Schmidt} S.~J.,  {Burgasser}
  A.~J.,  {Williams} P. K.~G.,   {Berger} E.,  2018, \mn@doi [\apj]
  {10.3847/1538-4357/aab8fe}, \href
  {https://ui.adsabs.harvard.edu/abs/2018ApJ...858...55P} {858, 55}

\bibitem[\protect\citeauthoryear{{Paudel}, {Gizis}, {Mullan}, {Schmidt},
  {Burgasser}  \& {Williams}}{{Paudel}
  et~al.}{2020}]{Paudel+etal+2020_LDwarf_flares}
{Paudel} R.~R.,  {Gizis} J.~E.,  {Mullan} D.~J.,  {Schmidt} S.~J.,  {Burgasser}
  A.~J.,   {Williams} P.~K.~G.,  2020, \mn@doi [\mnras]
  {10.1093/mnras/staa1137}, \href
  {https://ui.adsabs.harvard.edu/abs/2020MNRAS.494.5751P} {494, 5751}

\bibitem[\protect\citeauthoryear{{Pelisoli} \& {Vos}}{{Pelisoli} \&
  {Vos}}{2019}]{Pelisoli+Joris+2019+ELM}
{Pelisoli} I.,  {Vos} J.,  2019, \mn@doi [\mnras] {10.1093/mnras/stz1876},
  \href {https://ui.adsabs.harvard.edu/abs/2019MNRAS.488.2892P} {488, 2892}

\bibitem[\protect\citeauthoryear{{Pietrukowicz} et~al.,}{{Pietrukowicz}
  et~al.}{2017}]{Pietrukowicz+etal+2017+blap}
{Pietrukowicz} P.,  et~al., 2017, \mn@doi [Nature Astronomy]
  {10.1038/s41550-017-0166}, \href
  {https://ui.adsabs.harvard.edu/abs/2017NatAs...1E.166P} {1, 0166}

\bibitem[\protect\citeauthoryear{{Pojmanski}}{{Pojmanski}}{2002}]{Pojmanski+2002+06hvariable}
{Pojmanski} G.,  2002, \actaa, \href
  {https://ui.adsabs.harvard.edu/abs/2002AcA....52..397P} {52, 397}

\bibitem[\protect\citeauthoryear{{Ramsay} \& {Hakala}}{{Ramsay} \&
  {Hakala}}{2005}]{RATS+2005+survey}
{Ramsay} G.,  {Hakala} P.,  2005, \mn@doi [\mnras]
  {10.1111/j.1365-2966.2005.09035.x}, \href
  {https://ui.adsabs.harvard.edu/abs/2005MNRAS.360..314R} {360, 314}

\bibitem[\protect\citeauthoryear{{Rappaport}, {Verbunt}  \& {Joss}}{{Rappaport}
  et~al.}{1983}]{Rappaport+etal+1983}
{Rappaport} S.,  {Verbunt} F.,   {Joss} P.~C.,  1983, \mn@doi [\apj]
  {10.1086/161569}, \href
  {https://ui.adsabs.harvard.edu/abs/1983ApJ...275..713R} {275, 713}

\bibitem[\protect\citeauthoryear{{Ratzloff}, {Law}, {Fors}, {Corbett},
  {Howard}, {del Ser}  \& {Haislip}}{{Ratzloff}
  et~al.}{2019}]{Evryscope+2019+performance}
{Ratzloff} J.~K.,  {Law} N.~M.,  {Fors} O.,  {Corbett} H.~T.,  {Howard} W.~S.,
  {del Ser} D.,   {Haislip} J.,  2019, \mn@doi [\pasp]
  {10.1088/1538-3873/ab19d0}, \href
  {https://ui.adsabs.harvard.edu/abs/2019PASP..131g5001R} {131, 075001}

\bibitem[\protect\citeauthoryear{{Rau} et~al.,}{{Rau}
  et~al.}{2009}]{PTF+2009+OT}
{Rau} A.,  et~al., 2009, \mn@doi [\pasp] {10.1086/605911}, \href
  {https://ui.adsabs.harvard.edu/abs/2009PASP..121.1334R} {121, 1334}

\bibitem[\protect\citeauthoryear{{Ricker} et~al.,}{{Ricker}
  et~al.}{2014}]{Ricker+etal+2014+TESS}
{Ricker} G.~R.,  et~al., 2014, in {Oschmann} Jacobus~M. J.,  {Clampin} M.,
  {Fazio} G.~G.,   {MacEwen} H.~A.,  eds,  Society of Photo-Optical
  Instrumentation Engineers (SPIE) Conference Series Vol. 9143, Space
  Telescopes and Instrumentation 2014: Optical, Infrared, and Millimeter Wave.
  p. 914320 (\mn@eprint {arXiv} {1406.0151}), \mn@doi{10.1117/12.2063489}

\bibitem[\protect\citeauthoryear{{Ricker} et~al.,}{{Ricker}
  et~al.}{2015}]{TESS+2015}
{Ricker} G.~R.,  et~al., 2015, \mn@doi [Journal of Astronomical Telescopes,
  Instruments, and Systems] {10.1117/1.JATIS.1.1.014003}, \href
  {https://ui.adsabs.harvard.edu/abs/2015JATIS...1a4003R} {1, 014003}

\bibitem[\protect\citeauthoryear{{R{\"o}ser}, {Schilbach}, {Schwan},
  {Kharchenko}, {Piskunov}  \& {Scholz}}{{R{\"o}ser}
  et~al.}{2008}]{Roser+etal+2008+PPMX}
{R{\"o}ser} S.,  {Schilbach} E.,  {Schwan} H.,  {Kharchenko} N.~V.,  {Piskunov}
  A.~E.,   {Scholz} R.~D.,  2008, \mn@doi [\aap] {10.1051/0004-6361:200809775},
  \href {https://ui.adsabs.harvard.edu/abs/2008A&A...488..401R} {488, 401}

\bibitem[\protect\citeauthoryear{{Scargle}}{{Scargle}}{1982}]{Scargle+1982}
{Scargle} J.~D.,  1982, \mn@doi [\apj] {10.1086/160554}, \href
  {https://ui.adsabs.harvard.edu/abs/1982ApJ...263..835S} {263, 835}

\bibitem[\protect\citeauthoryear{{Shah}, {van der Sluys}  \& {Nelemans}}{{Shah}
  et~al.}{2012}]{Shah+etal+2012}
{Shah} S.,  {van der Sluys} M.,   {Nelemans} G.,  2012, \mn@doi [\aap]
  {10.1051/0004-6361/201219309}, \href
  {https://ui.adsabs.harvard.edu/abs/2012A&A...544A.153S} {544, A153}

\bibitem[\protect\citeauthoryear{{Shin}, {Sekora}  \& {Byun}}{{Shin}
  et~al.}{2009}]{Shin+etal+2009+variable}
{Shin} M.-S.,  {Sekora} M.,   {Byun} Y.-I.,  2009, \mn@doi [\mnras]
  {10.1111/j.1365-2966.2009.15576.x}, \href
  {https://ui.adsabs.harvard.edu/abs/2009MNRAS.400.1897S} {400, 1897}

\bibitem[\protect\citeauthoryear{{Sokolovsky} et~al.,}{{Sokolovsky}
  et~al.}{2017}]{Sokolovsky+etal+2017+variables}
{Sokolovsky} K.~V.,  et~al., 2017, \mn@doi [\mnras] {10.1093/mnras/stw2262},
  \href {https://ui.adsabs.harvard.edu/abs/2017MNRAS.464..274S} {464, 274}

\bibitem[\protect\citeauthoryear{{Tamuz}, {Mazeh}  \& {Zucker}}{{Tamuz}
  et~al.}{2005}]{Tamuz+etal+2005+sysremove}
{Tamuz} O.,  {Mazeh} T.,   {Zucker} S.,  2005, \mn@doi [\mnras]
  {10.1111/j.1365-2966.2004.08585.x}, \href
  {https://ui.adsabs.harvard.edu/abs/2005MNRAS.356.1466T} {356, 1466}

\bibitem[\protect\citeauthoryear{{Thompson} et~al.,}{{Thompson}
  et~al.}{2019}]{Thompson+etal+2019+Science}
{Thompson} T.~A.,  et~al., 2019, \mn@doi [Science] {10.1126/science.aau4005},
  \href {https://ui.adsabs.harvard.edu/abs/2019Sci...366..637T} {366, 637}

\bibitem[\protect\citeauthoryear{{Tomaney} \& {Crotts}}{{Tomaney} \&
  {Crotts}}{1996}]{Tomaney+Crotts+1996+DIA}
{Tomaney} A.~B.,  {Crotts} A. P.~S.,  1996, \mn@doi [\aj] {10.1086/118228},
  \href {https://ui.adsabs.harvard.edu/abs/1996AJ....112.2872T} {112, 2872}

\bibitem[\protect\citeauthoryear{{Toonen}, {Voss}  \& {Knigge}}{{Toonen}
  et~al.}{2014}]{Toonen+etal+2014}
{Toonen} S.,  {Voss} R.,   {Knigge} C.,  2014, \mn@doi [\mnras]
  {10.1093/mnras/stu569}, \href
  {https://ui.adsabs.harvard.edu/abs/2014MNRAS.441..354T} {441, 354}

\bibitem[\protect\citeauthoryear{{VanderPlas}}{{VanderPlas}}{2018}]{VanderPlas+2018+LSP_understanding}
{VanderPlas} J.~T.,  2018, \mn@doi [\apjs] {10.3847/1538-4365/aab766}, \href
  {https://ui.adsabs.harvard.edu/abs/2018ApJS..236...16V} {236, 16}

\bibitem[\protect\citeauthoryear{{Walkowicz} et~al.,}{{Walkowicz}
  et~al.}{2011}]{Walkowicz+etal+2011+flares}
{Walkowicz} L.~M.,  et~al., 2011, \mn@doi [\aj] {10.1088/0004-6256/141/2/50},
  \href {https://ui.adsabs.harvard.edu/abs/2011AJ....141...50W} {141, 50}

\bibitem[\protect\citeauthoryear{{Wang} et~al.,}{{Wang}
  et~al.}{2019}]{Wang+etal+2019+RV_measurements}
{Wang} R.,  et~al., 2019, \mn@doi [\apjs] {10.3847/1538-4365/ab3cc0}, \href
  {https://ui.adsabs.harvard.edu/abs/2019ApJS..244...27W} {244, 27}

\bibitem[\protect\citeauthoryear{{Wang}, {Chen}, {Liu}, {Chen}, {Wu}, {Tang},
  {Guo}  \& {Han}}{{Wang} et~al.}{2021}]{Wang+etal+2021}
{Wang} B.,  {Chen} W.,  {Liu} D.,  {Chen} H.,  {Wu} C.,  {Tang} W.,  {Guo} Y.,
   {Han} Z.,  2021, arXiv e-prints, \href
  {https://ui.adsabs.harvard.edu/abs/2021arXiv210601369W} {p. arXiv:2106.01369}

\bibitem[\protect\citeauthoryear{{Watson}, {Henden}  \& {Price}}{{Watson}
  et~al.}{2006}]{Watson+etal+2006+VSX}
{Watson} C.~L.,  {Henden} A.~A.,   {Price} A.,  2006, Society for Astronomical
  Sciences Annual Symposium, \href
  {https://ui.adsabs.harvard.edu/abs/2006SASS...25...47W} {25, 47}

\bibitem[\protect\citeauthoryear{{Yang} \& {Liu}}{{Yang} \&
  {Liu}}{2019}]{Yang+Liu+2019+flare_catalog}
{Yang} H.,  {Liu} J.,  2019, \mn@doi [\apjs] {10.3847/1538-4365/ab0d28}, \href
  {https://ui.adsabs.harvard.edu/abs/2019ApJS..241...29Y} {241, 29}

\bibitem[\protect\citeauthoryear{{Yang}, {Liu}, {Qiao}, {Zhang}, {Gao}, {Cui}
  \& {Han}}{{Yang} et~al.}{2018}]{Yang+etal+2018+SC_LC_flares}
{Yang} H.,  {Liu} J.,  {Qiao} E.,  {Zhang} H.,  {Gao} Q.,  {Cui} K.,   {Han}
  H.,  2018, \mn@doi [\apj] {10.3847/1538-4357/aabd31}, \href
  {https://ui.adsabs.harvard.edu/abs/2018ApJ...859...87Y} {859, 87}

\bibitem[\protect\citeauthoryear{{Yang} et~al.,}{{Yang}
  et~al.}{2020}]{Yang+etal+2020+EB_lamost}
{Yang} F.,  et~al., 2020, \mn@doi [\apjs] {10.3847/1538-4365/ab9b77}, \href
  {https://ui.adsabs.harvard.edu/abs/2020ApJS..249...31Y} {249, 31}

\bibitem[\protect\citeauthoryear{{Yi}, {Sun}  \& {Gu}}{{Yi}
  et~al.}{2019}]{Yi+etal+2019+theory_bh}
{Yi} T.,  {Sun} M.,   {Gu} W.-M.,  2019, \mn@doi [\apj]
  {10.3847/1538-4357/ab4a75}, \href
  {https://ui.adsabs.harvard.edu/abs/2019ApJ...886...97Y} {886, 97}

\bibitem[\protect\citeauthoryear{{Zhang} \& {Bloom}}{{Zhang} \&
  {Bloom}}{2021}]{Zhang+Bloom+2021+nn}
{Zhang} K.,  {Bloom} J.~S.,  2021, \mn@doi [\mnras] {10.1093/mnras/stab1248},
  \href {https://ui.adsabs.harvard.edu/abs/2021MNRAS.505..515Z} {505, 515}

\bibitem[\protect\citeauthoryear{{Zhang}, {Ge}, {Lu}, {Cao}, {Chen}, {Mao}  \&
  {Jiang}}{{Zhang} et~al.}{2015}]{Zhang+etal+2015+Xinglong_conditions}
{Zhang} J.-C.,  {Ge} L.,  {Lu} X.-M.,  {Cao} Z.-H.,  {Chen} X.,  {Mao} Y.-N.,
  {Jiang} X.-J.,  2015, \mn@doi [\pasp] {10.1086/684369}, \href
  {https://ui.adsabs.harvard.edu/abs/2015PASP..127.1292Z} {127, 1292}

\bibitem[\protect\citeauthoryear{{Zhang} et~al.,}{{Zhang}
  et~al.}{2020}]{TMTS+2020+survey}
{Zhang} J.-C.,  et~al., 2020, \mn@doi [\pasp] {10.1088/1538-3873/abbea2}, \href
  {https://ui.adsabs.harvard.edu/abs/2020PASP..132l5001Z} {132, 125001}

\bibitem[\protect\citeauthoryear{{Zhao}, {Zhao}, {Chu}, {Jing}  \&
  {Deng}}{{Zhao} et~al.}{2012}]{Zhao+etal+2012+LAMOST}
{Zhao} G.,  {Zhao} Y.-H.,  {Chu} Y.-Q.,  {Jing} Y.-P.,   {Deng} L.-C.,  2012,
  \mn@doi [Research in Astronomy and Astrophysics]
  {10.1088/1674-4527/12/7/002}, \href
  {https://ui.adsabs.harvard.edu/abs/2012RAA....12..723Z} {12, 723}

\bibitem[\protect\citeauthoryear{{Zheng} et~al.,}{{Zheng}
  et~al.}{2019}]{Zheng+etal+2019+lamost_bh}
{Zheng} L.-L.,  et~al., 2019, \mn@doi [\aj] {10.3847/1538-3881/ab449f}, \href
  {https://ui.adsabs.harvard.edu/abs/2019AJ....158..179Z} {158, 179}

\bibitem[\protect\citeauthoryear{{Zhu}, {L{\"u}}  \& {Wang}}{{Zhu}
  et~al.}{2012}]{Zhu+etal+2012}
{Zhu} C.-H.,  {L{\"u}} G.-L.,   {Wang} Z.-J.,  2012, \mn@doi [Research in
  Astronomy and Astrophysics] {10.1088/1674-4527/12/11/007}, \href
  {https://ui.adsabs.harvard.edu/abs/2012RAA....12.1526Z} {12, 1526}

\makeatother
\end{thebibliography}
\input{TMTS.bbl}




\bsp	
\label{lastpage}
\end{document}